\documentclass{emulateapj}

\usepackage{graphics}

\newcommand{\msun}{M_\odot}
\newcommand{\mstar}{${M_{\rm stellar}}$}
\newcommand{\gyr}{{\rm Gyr}}
\newcommand{\myr}{{\rm Myr}}
\newcommand{\mum}{$\mu$m}

\newcommand{\zdist}[3]%
{
\begin{#1}[#2]
\begin{centering}
  \includegraphics[width=#3\textwidth]{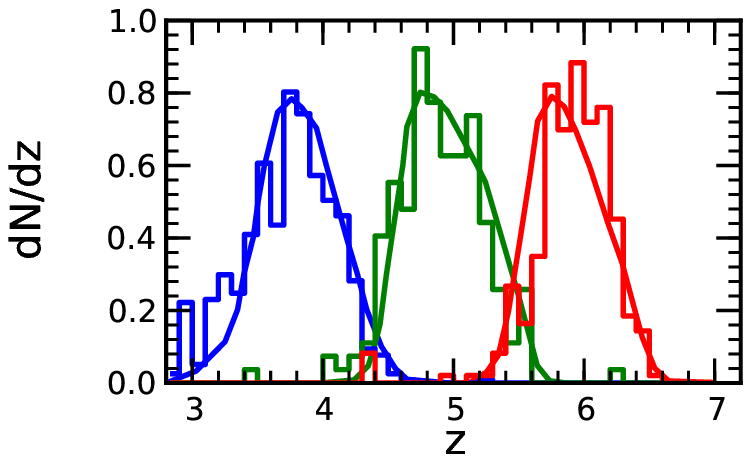}
  \caption{The photometric redshift distribution of each of the samples
    (histograms) compared to the redshift distribution expected from the
    color selection criteria described in Section 2.1 (solid lines). The
    photometric redshifts are determined using the rest-frame UV
    phtometry only and ignoring the IRAC photometry to avoid possible
    effects introduced by the rest-frame optical emission lines on the
    photometric redshifts. The redshifts are primarily driven by the
    observed wavelength of the Lyman Break. The photometric redshifts
    shown here are used for the SED modeling throughout the analysis.
  }
  \label{fig:zdist}
\end{centering}
\end{#1}
}

\newcommand{\colormagB}[3]%
{
\begin{#1}[#2]
\begin{centering}
  \includegraphics[width=#3\textwidth]{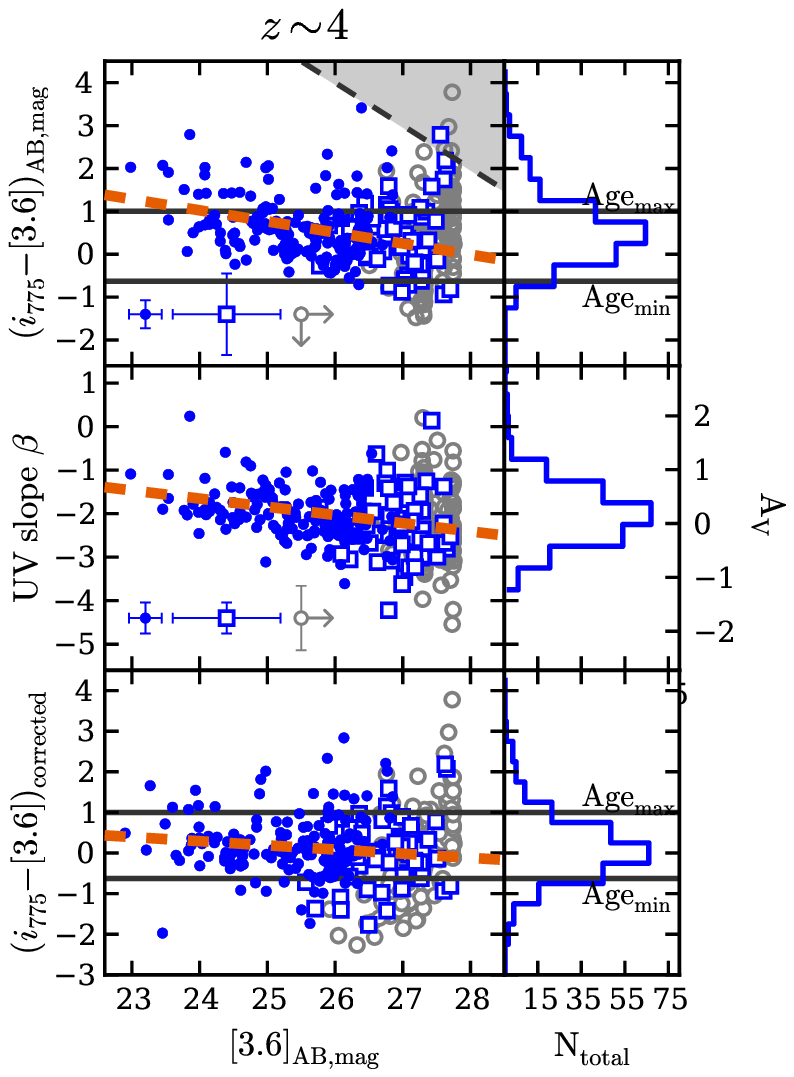}
  \caption{
    The observed colors of $z\sim4$ LBGs. Only sources with reliable
    IRAC photometry are included (see text, section 2.2).  In all panels
    the solid blue circles correspond to sources that are detected in
    IRAC [3.6] ($>2\sigma$), open squares are used for marginally
    detected sources ($1\mbox{--}2\sigma$), and gray open symbols are
    for IRAC-undetected ($<1\sigma$) sources (plotted as $1\,\sigma$
    limits).The histograms in the right panels show the color
    distributions. Top: the $i_{775}-[3.6]$ color as a function of
    magnitude in the [3.6] channel. The dashed line in the top right
    corner corresponds to the limiting magnitudes probed by our
    \emph{HST} observations. The three symbols in the lower left show
    the typical uncertainties. The solid horizontal lines show the
    minimum and maximum color for a dust-free stellar population with a
    CSF history and ages between 10 \myr~and the age of the universe at
    $z=3.5$. The thick dashed line is a fit to the detected sources and
    shows a trend of redder colors for brighter galaxies. The best fit
    corresponds to $(i_{775}-[3.6])=-0.25 \times [3.6]+{\textit const}$.
    Middle: the UV slope $\beta$ ($f_\lambda\propto\lambda^\beta$)
    determined from a linear fit to the available \emph{HST} photometry
    of each source (e.g., as done by \citealt{bouw12, cast12}). This
    color also shows a trend of redder colors for brighter sources that
    is consistent with previous reports \citep[e.g.,][]{bouw12}. The
    best fit line corresponds to $\beta=-0.19\times[3.6]+2.87$. In our
    stellar population modeling we assume that this trend is caused
    exclusively by a varying amount of dust reddening (\citealt{meur99}
    relation, Equation \ref{eq:meurer}).  Bottom : the $i_{775}-[3.6]$
    color after dust reddening corrections are applied. These
    corrections are derived from the UV slope. The trend of redder
    colors for brighter sources is weaker, with a best fit corresponding
    to $(i_{775}-[3.6])_{\rm dust-corrected}=-0.10 \times [3.6]+{\textit
    const}$. After this correction, most sources lie within the range of
    colors expected for dust-free CSF models.  The residual trend could
    be caused by a weak trend of age versus luminosity.
  }
  \label{fig:colormagB}
\end{centering}
\end{#1}
}

\newcommand{\colormagRest}[3]%
{
\begin{#1}[#2]
\begin{centering}
  \includegraphics[width=#3\textwidth]{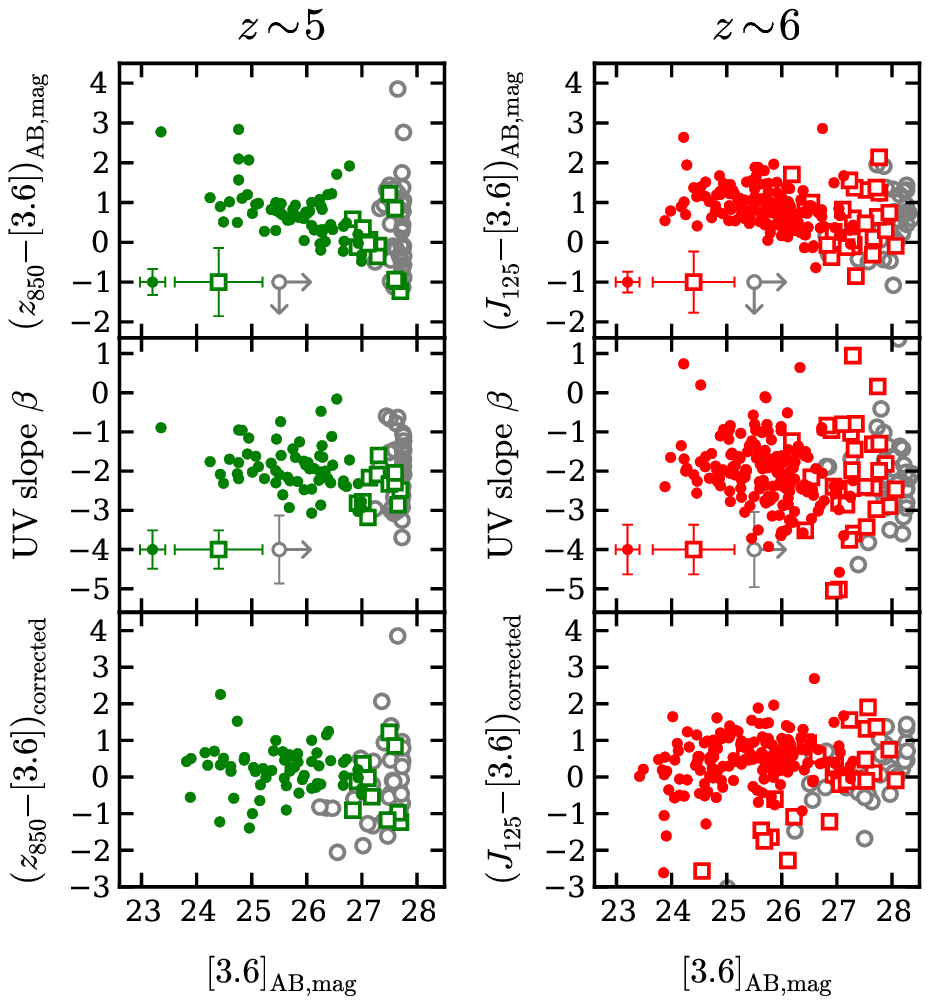}
  \caption{The observed colors of $z\sim5$ and 6 LBGs. \emph{Left}
  column (green points) show the colors of the $V-$dropouts
  ($z\sim5$).  \emph{Right} column (red symbols) shows the colors of
  the $i-$dropouts ($z\sim6$). Symbols are as in Figure
  \ref{fig:colormagB}. The same trends of bluer colors (both the
  UV-slope and the UV-to-optical colors) for fainter sources are
  present at these redshifts.}.
  \label{fig:colormagRest}
\end{centering}
\end{#1}
}

\newcommand{\sfrvsmstar}[3]%
{
\begin{#1}[#2]
\begin{centering}
  \includegraphics[width=#3\textwidth]{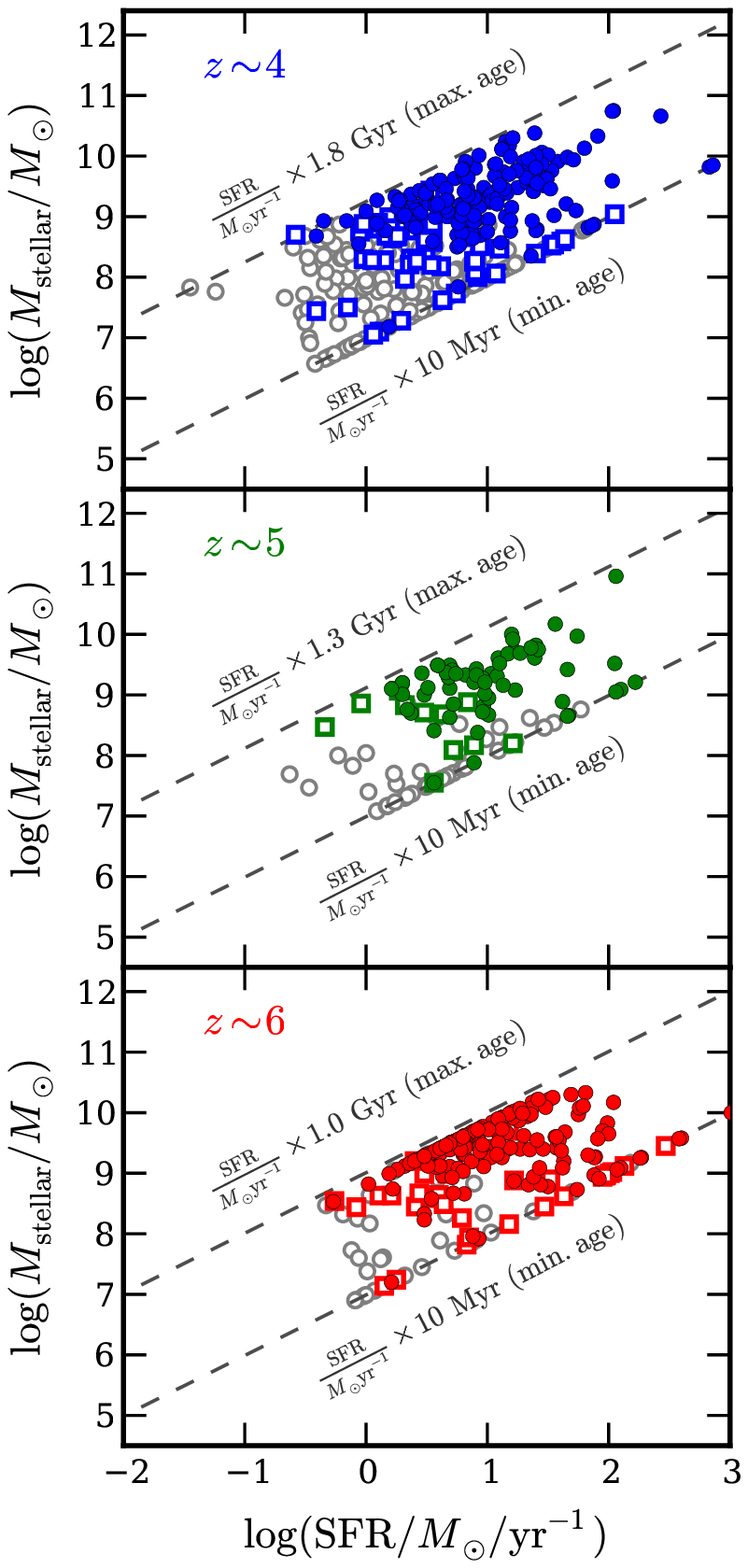}
  \caption{The SFR versus \mstar~relation for for our samples at
  $z\sim4$ (top, blue), $z\sim5$ (middle, green), and $z\sim6$ (botom,
  red).  Only sources with reliable cleaned IRAC photometry are
  considered (see text Section 2.2). Stellar masses and SFRs were
  derived assuming a CSF history. The dust reddening in the model has
  been derived directly from the UV slopes $\beta$ of each sources
  according to Equation \ref{eq:meurer}. In all panels, solid circles
  indicate sources that are detected ($>2\sigma$) in the
  $Spitzer$/IRAC [3.6] channel and hence their masses can be more
  accurately estimated; open squares indicate marginal detections
  ($1-2\sigma$); and open gray circles indicate IRAC undetected
  galaxies. The dashed lines indicate the minimum and maximum
  \mstar~that a galaxy can reach assuming that it has been forming
  stars at a constant rate, for the minimum age included in our models
  (10 \myr) and for the age of the universe at $z\sim3.5, 4.5$, and
  5.5 (from top to bottom).}
  \label{fig:sfrvsmstar}
\end{centering}
\end{#1}
}

\newcommand{\SFRvsSFRnew}[3]%
{
\begin{#1}[#2]
\begin{centering}
  \includegraphics[width=#3\textwidth]{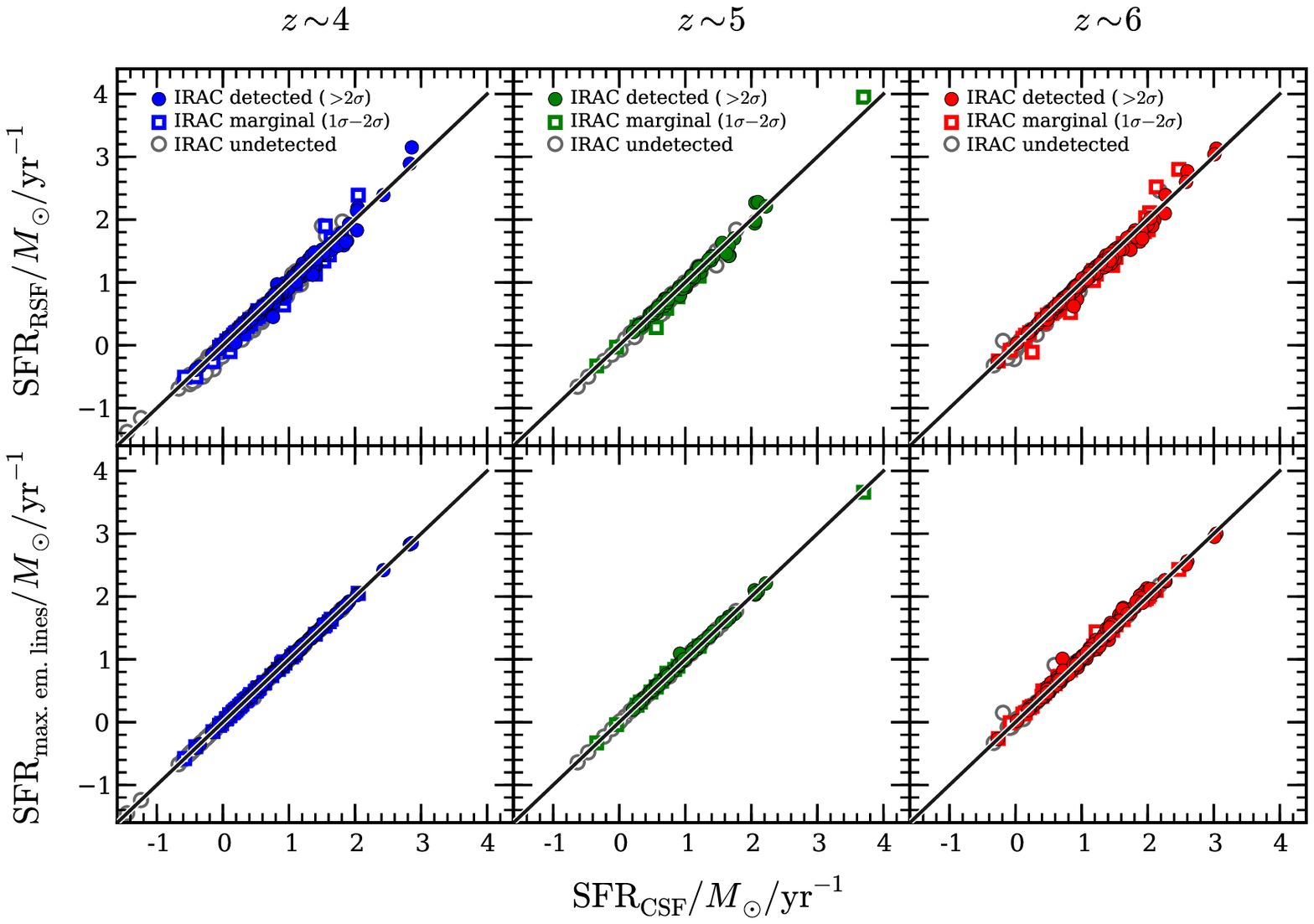}
  \caption{Comparison of the SFRs derived using different model
    assumptions. In all panels, the solid symbols represent sources that
    are detected ($>2\sigma$) in IRAC [3.6], open squares show marginal
    detections ($1\mbox{--}2\sigma$), and open gray circles show IRAC
    undetected sources. The top row shows the comparison of the SFRs
    derived assuming rising star formation histories and constant star
    formation histories. The bottom row shows the effect that correcting
    for optical emission lines have on the SFRs. The solid lines
    corresponds to the identity. Overall, we find that the derived SFRs
    are not very sensitive to the model assumptions.
  }
  \label{fig:sfrvssfr}
\end{centering}
\end{#1}
}

\newcommand{\mstarmstar}[3]%
{
\begin{#1}[#2]
\begin{centering}
  \includegraphics[width=#3\textwidth]{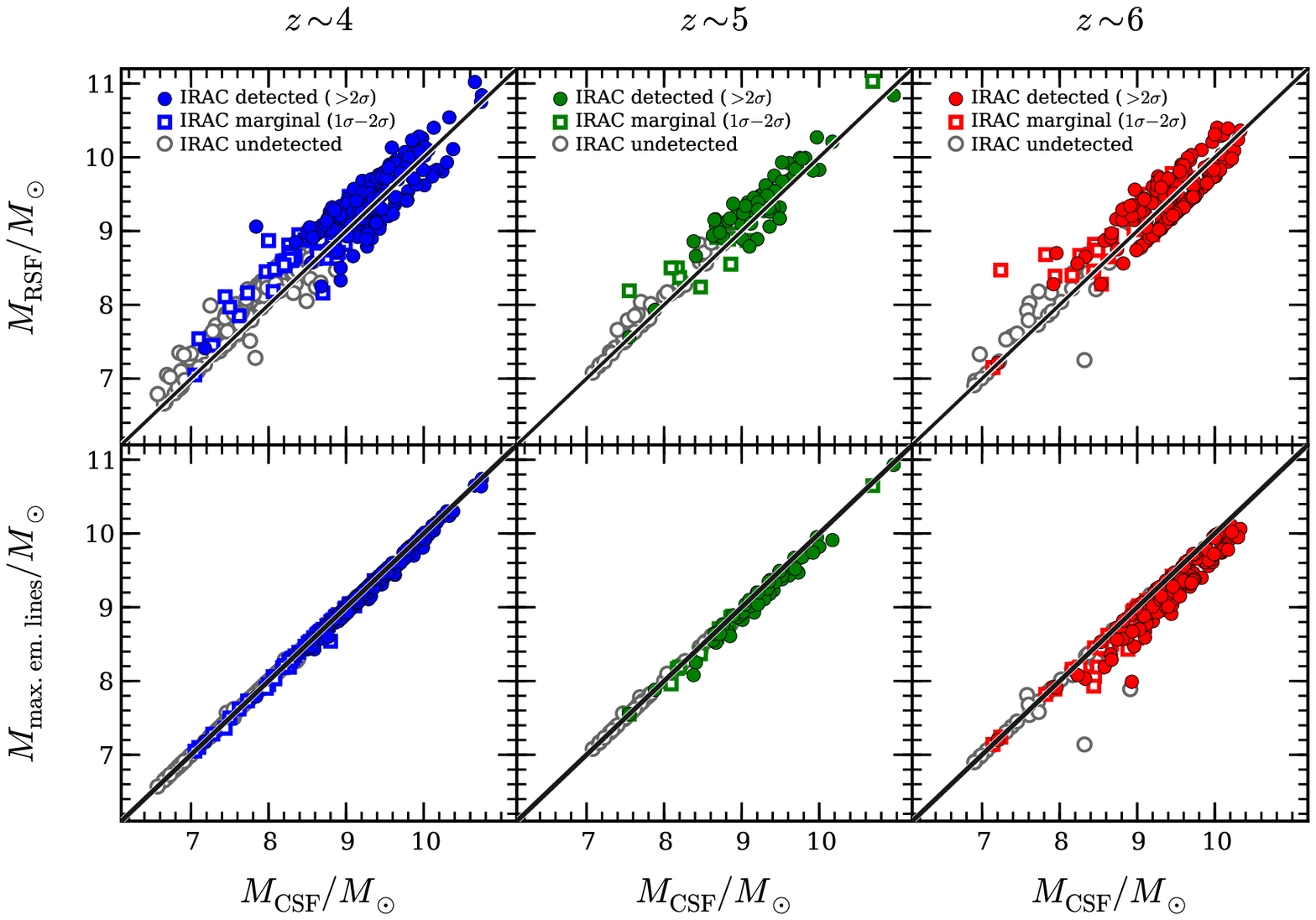}
  \caption{Comparisons of the stellar masses we derive using different
    model assumptions. In all panels, the solid symbols represent
    sources that are detected ($>2\sigma$) in IRAC [3.6], open squares
    show marginal detections ($1\mbox{--}2\sigma$), and open gray
    circles show IRAC undetected sources. The solid line corresponds to
    the identity. The left column shows the $z\sim4$ sample (blue), the
    middle is the $z\sim5$ (green), and the right column shows the
    $z\sim6$ sample. The horizontal axis shows the stellar masses
    determined from our CSF model. The vertical axis for the upper set
    of panels shows the stellar masses derived from a model with
    exponentially rising star formation rate (${\rm SFR}\propto
    e^{(t-t_0)/\tau}$, where $t-t_0$ is the age, and $\tau=500~\myr$).
    This model is explicitly constructed to reproduce the observed
    evolution of the UV LF. Both $M_{\rm CSF}$ and $M_{\rm RSF}$ are
    derived with models with stellar continuum only. When only the
    detected sources are considered, the mean $M_{\rm CSF}$ and the
    $M_{\rm RSF}$ are consistent with each other, but with a scatter of
    $\sim0.25\,$dex. The IRAC undetected and marginal detections show a
    slightly larger mass for the RSF determinations. The vertical axis
    for the lower set of panels shows the determinations of the stellar
    mass using a model that assumes strong optical emission lines which
    evolve with redshift (see Section 3.4).  A CSF is assumed. The
    strength of the emission lines assumed can be seen in Figure
    \ref{fig:emlineCorrections}. The effects on the stellar mass are
    larger at $z\sim6$ where they cause an average decrease in the
    stellar mass of $\sim0.26~$dex for IRAC detected sources. This is
    expected as both IRAC channels present large contributions of flux
    from emission lines at this redshift.
  }
  \label{fig:mstarmstar}
\end{centering}
\end{#1}
}

\newcommand{\emlineCorrections}[3]%
{
\begin{#1}[#2]
\begin{centering}
  \includegraphics[width=#3\textwidth]{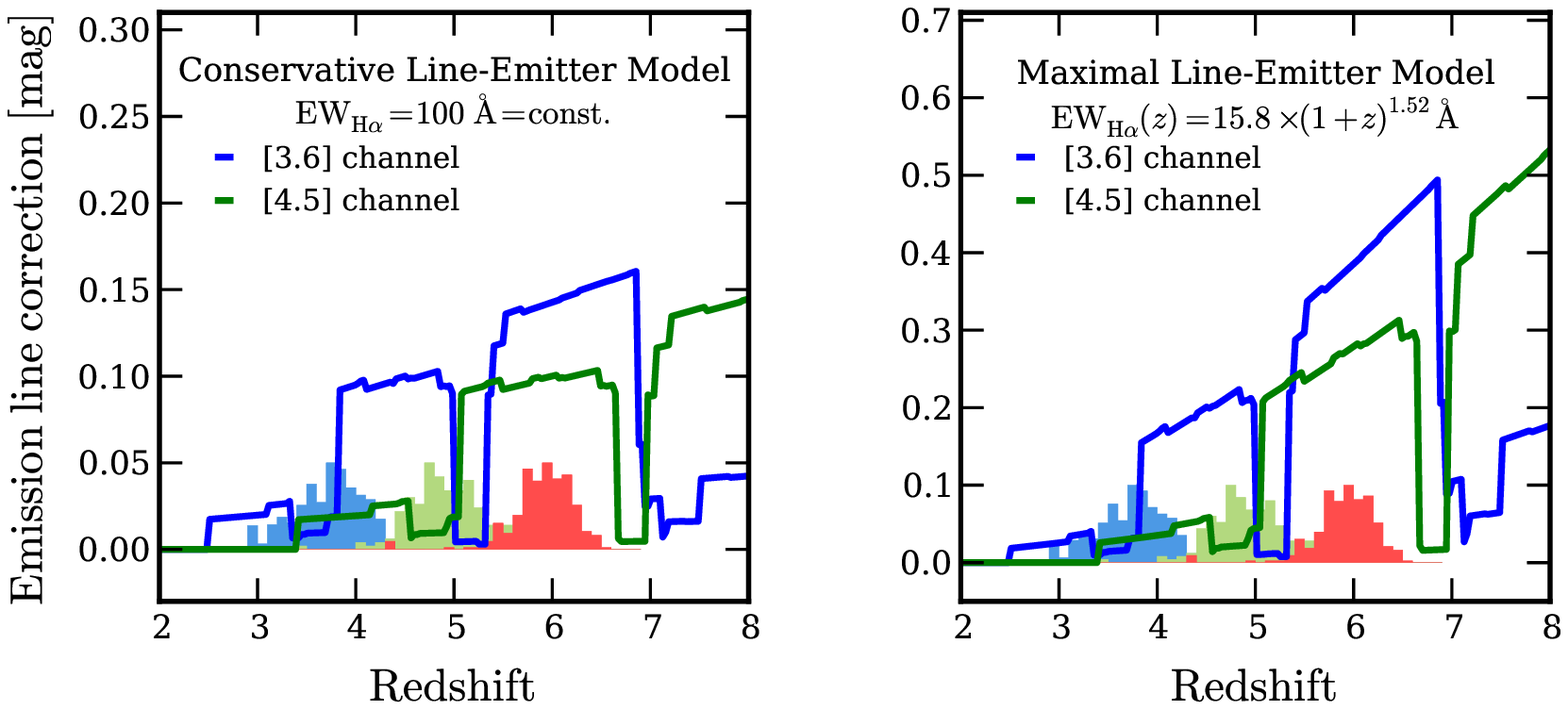}
  \caption{The possible effect of strong rest-frame optical emission
  lines in the $Spitzer$/IRAC photometry. In both panels the blue
  curves show the corrections, in magnitudes, that need to be applied
  to the [3.6] photometry to remove the flux from emission lines. The
  green curve is the equivalent for the [4.5] photometry. In the
  \emph{left} panel we conservatively assume a constant H$\alpha$
  rest-frame EW of 100 \AA. To estimate the strength of the other
  lines we assume the flux ratios presented by \citet{ande03}, for a
  metallicity $Z=0.2\,Z_\odot$. We also assume a flat underlying
  continuum (in F$_\nu$). In the \emph{right} panel the H$\alpha$ EW
  is estimated from the trends observed as a function of redshift for
  sources with $\log_{10}(M_{\rm stellar}/M_\odot)=10\mbox{--}10.5$
  \citep{fuma12}.  This extrapolation is based on recent \emph{HST}
  grism observations and predicts H$\alpha$ EW$_0\sim300\,$\AA\ at
  $z\sim6$.  The light blue, light green, and light red, solid
  histograms (arbitrary normalization), show the redshift distribution
  of our sample. These corrections result in lower stellar masses but,
  in general, they do not affect the SFRs.}
  \label{fig:emlineCorrections}
\end{centering}
\end{#1}
}

\newcommand{\highzSSFR}[3]%
{
\begin{#1}[#2]
\begin{centering}
  \includegraphics[width=#3\textwidth]{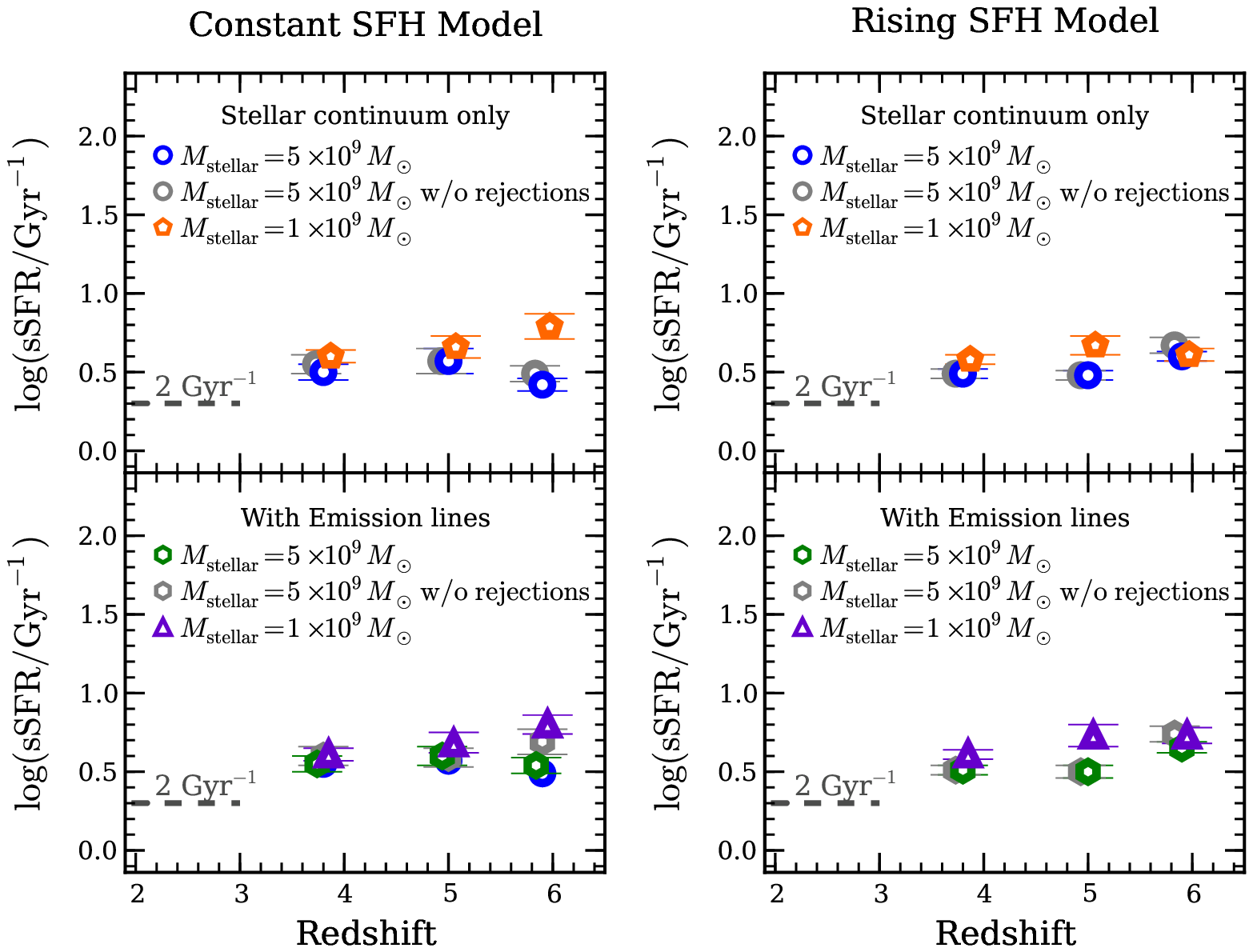}
  \caption{The mean specific SFR as a function of redshift for our
    samples at $z\sim4,~5,$ and 6. The mean sSFR was determined for
    galaxies in two different mass bins $1\times10^9\ M_\odot$ and
    $5\times10^9\ M_\odot$. Top-left: the mean sSFR derived assuming a
    CSF history and models with stellar continuum only (i.e., ignoring
    the effects of emission lines). The mean values were estimated after
    removing extreme values (${\rm sSFR \sim 100~Gyr^{-1}}$) which
    correspond to the minimum ages in our grid of models (10 Myr). This
    rejection does not cause significant differences, as can be seen by
    comparing the blue (with rejection) and gray circles (without
    rejection). Top-right: mean sSFR derived assuming an exponentially
    rising SFH and models with stellar continuum only. The mean sSFR
    values do not change significantly for the RSF model relative to
    that for a CSF model (see Top-left panel). Bottom-left: assuming our
    \emph{maximal} model for the average strength of the emission lines,
    we re-model the galaxies first assuming a CSF SFH. The behavior is
    very similar to that shown in the top-left panel -- blue open
    circles are repeated here for comparison. Rejecting extreme values
    makes a slightly larger difference in this case but only at
    $z\sim6$.  Bottom-right: the mean sSFR derived assuming an
    exponentially rising SFH and also assuming our \emph{maximal} model
    for the average strength of the emission lines -- this is our
    preferred model. Again very similar behavior is seen. Overall, the
    mean sSFR at $z\sim4$ and 5 varies weakly with both redshift and
    mass in all models. The data suggest that galaxies in the lower mass
    bin may have slightly larger mean sSFR values but the evidence is
    weak and the difference is at most 0.1 dex. Differences appear at
    $z\sim6$ but depend on the model assumptions. In all cases, the sSFR
    evolution with redshift is only mild.
  }
  \label{fig:SSFRcomparison}
\end{centering}
\end{#1}
}

\newcommand{\muvmstar}[3]%
{
\begin{#1}[#2]
\begin{centering}
  \includegraphics[width=#3\textwidth]{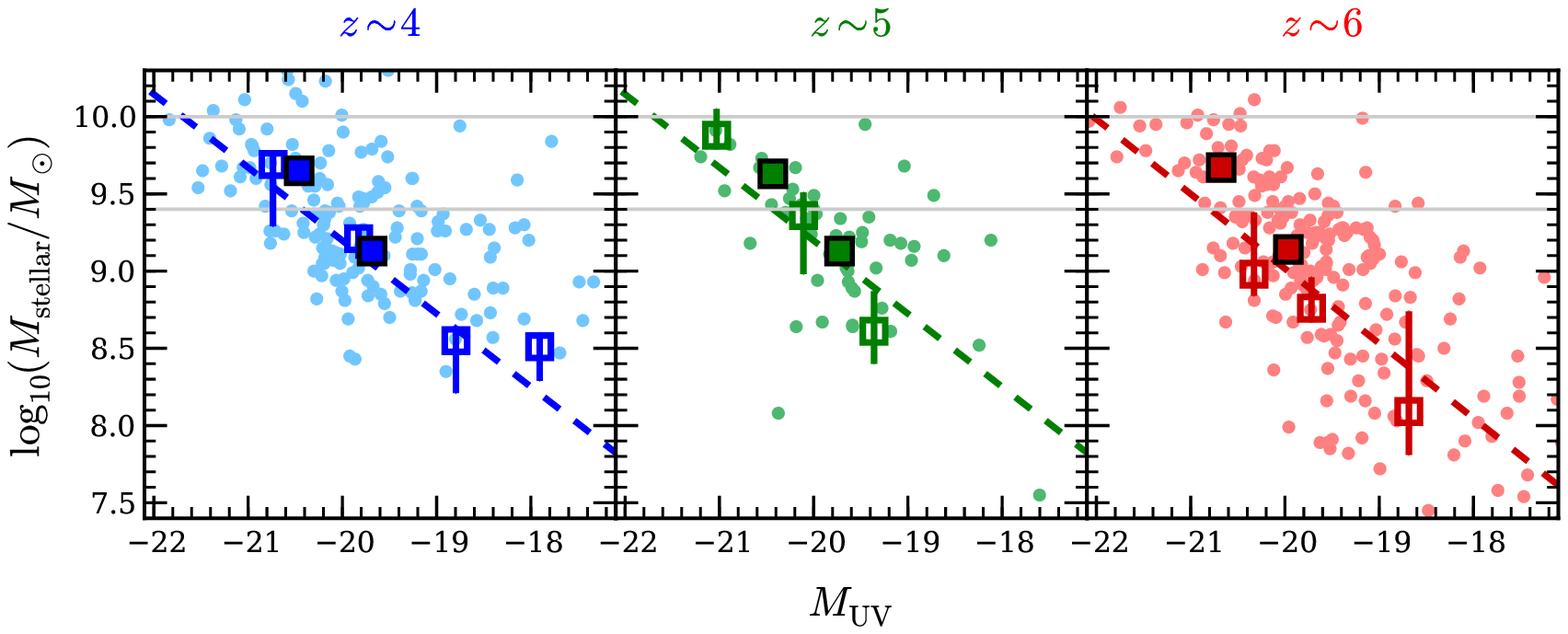}
  \caption{$\log(M_{\rm stellar})$ versus $M_{\rm UV}$ for the galaxies
    in our samples. Only individual glaxies that are detected in IRAC
    ($>2\,\sigma$) are plotted as the background points. The large open
    squares corresond to the stacked SEDs presented in \citet{gonz11}
    which are binned by UV luminosity. The modeling of the stacked SEDs
    includes corrections for emission lines as in our \emph{maximal}
    model. The dashed line is the best fit trend assumed in
    \citet{star13} after their modeling of the same stacked SEDs using
    similar assumptions. Our estimates (open squares) are in good
    agreement with their derived trend (dashed line). The solid squares
    show the results of binning the individual sources by stellar mass
    (the high mass bin centered at $5\times10^9\,M_\odot$ is highlighted
    by the horizontal lines). Low-luminosity galaxies with high $M/L$
    ratios are more likely included in the bin than high-luminosity ones
    with low $M/L$ (simply because faint galaxies are more numerous).
    As a result, the average luminosity within a mass bin is biased low,
    resulting in lower a sSFR.
  } 
  \label{fig:muvmstar}
\end{centering}
\end{#1}
}

\newcommand{\sSFR}[3]%
{
\begin{#1}[#2]
\begin{centering}
  \includegraphics[width=#3\textwidth]{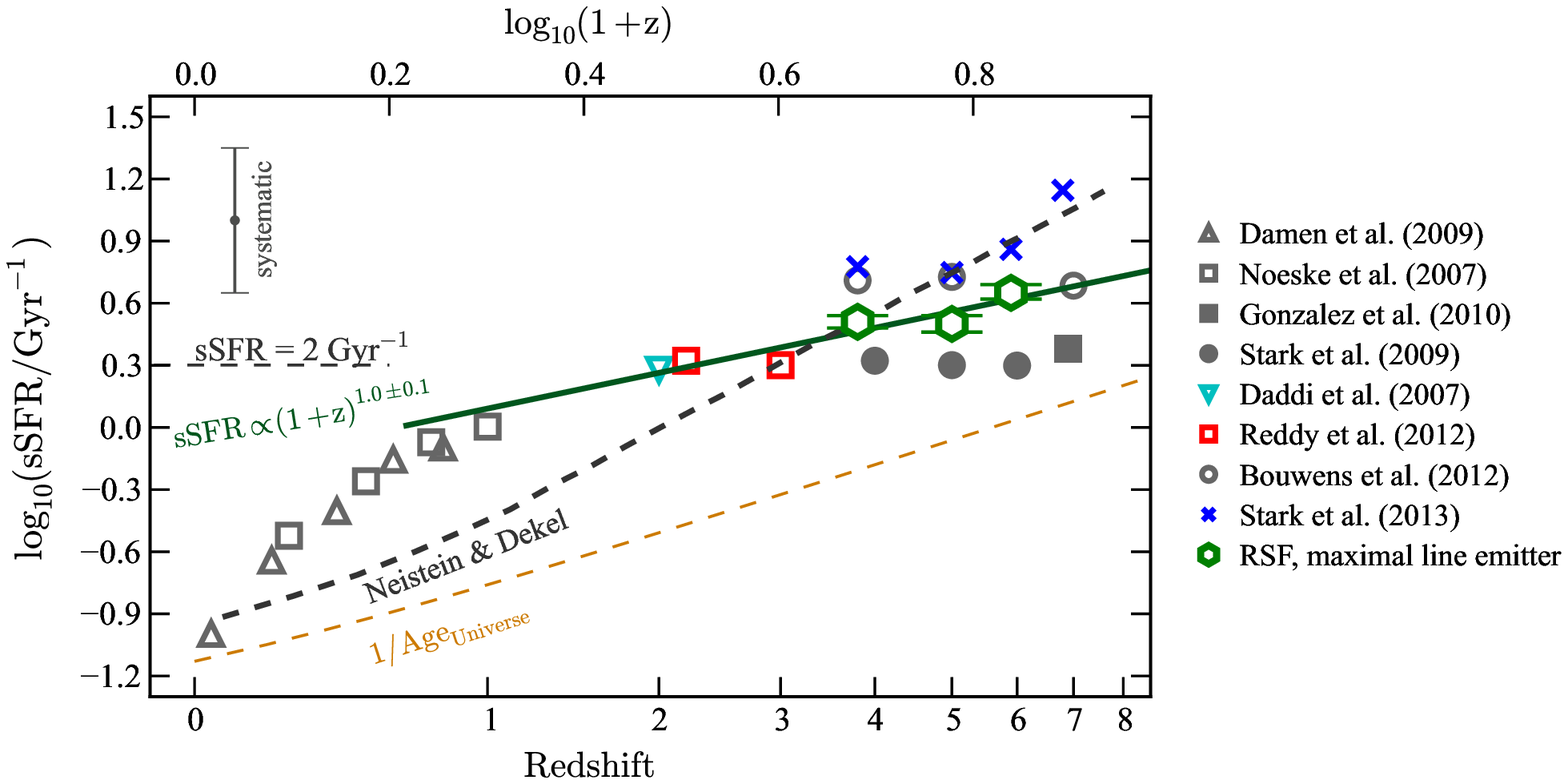}
  \caption{The mean specific SFR as a function of redshift for galaxies
    with estimated stellar masses $\log_{10}(M_{\rm stellar}/M_\odot) =
    9.4\mbox{--}10$, corresponding to our $5\times10^9~M_\odot$ bin. The
    values at $z<4$ are taken from the literature \citep{dame09, noes07,
    dadd07, redd12}. The solid gray points correspond to previous
    reports from the literature at $z\gtrsim4$ \citep{star09, gonz10}
    and the open gray points correspond to these values after a simple
    correction for dust extinction is made \citep{bouw12}. The blue
    crosses are from \citet{star13}.  The the open hexagons at
    $z\gtrsim4$ are the new results from our stellar population modeling
    of the full rest-frame UV and optical SED. These results were
    derived assuming an exponentially rising SFH (RSF) and the average
    emission line flux is subtracted based on the \emph{maximal} model
    described in Section 3.4. The dust reddening was derived using the
    \citet{meur99} relation and the measured UV slopes, $\beta$. The
    error bars correspond to random errors. Our estimate of the
    systematic error, is given as well in the upper left corner. Our
    sSFR estimates depend weakly on the modeling assumptions (see Figure
    \ref{fig:SSFRcomparison}). The differences between our results and
    those of \citet{star13} arise from them using UV-luminosity-binned
    averages versus our mass-binned averages (see Section 5.2 and the
    Appendix for details). A weighted best fit of our measured sSFR as a
    function of redshift at $z>2$ is: ${\rm \log(sSFR(z)) = -0.1(\pm0.1)
    + 1.0(\pm0.1) \log(1+z)}$ (which incorporates the values derived at
    $z\sim2$ by \citealt{dadd07}, $z\sim2\mbox{--}3$ by
    \citealt{redd12}, and the $z\gtrsim4$ from this work). The derived
    dependence of the sSFR on redshift is much weaker than that observed
    at $z<2$ and that expected from the theoretical expectations at
    $z>2$ (e.g., \citealt{neis08}, dashed line).
  }
  \label{fig:sSFR}
\end{centering}
\end{#1}
}

\newcommand{\tableSample}[1]%
{
\begin{deluxetable}{lcccc}[#1]
\tablecaption{\label{tbl:sampleSummary}Sample Summary.}
\tablehead{
  & \colhead{$z\sim4$} & \colhead{$z\sim5$} 
  & \colhead{$z\sim6$}
}
\startdata
ERS    &  270 (524)  &  77 (123)   &  29 (36)\\
HUDF   &  137 (205)  &  41 (55)    &  54 (59)\\
CANDELS &          &             &  143 (182)\\
\tableline\\
TOTAL  &  407 (729) &  118 (178)  &  226 (277)\\
\enddata
\tablecomments{
  Number of sources in our $z\sim4$, $z\sim5$, and $z\sim6$ samples.
  IRAC photometry of these sources requires fitting and subtraction of
  the flux from surrounding foreground neighbors.  This is not possible
  in all the cases. The table shows the number of sources in our samples
  with clean IRAC photometry and the total number of sources in these
  samples in parenthesis. To improve the number statistics at $z\sim6$
  we have extended the sample by including the GOODS-S CANDELS data.
}
\end{deluxetable}
}

\newcommand{\tableSSFR}[1]%
{
\begin{deluxetable*}{lccccccccc}[#1]
  \tablecaption{\label{tbl:SSFRresults} Current Estimates of the mean
  Specific Star Formation Rate for $z\sim4\mbox{--}6$ Galaxies}
  \tablehead{ & 
  \multicolumn{3}{c}{$z\sim4$} &
  \multicolumn{3}{c}{$z\sim5$} &
  \multicolumn{3}{c}{$z\sim6$} \\[0.4em]
  \colhead{Model} & 
  \colhead{$N_{\rm bin}$} & \colhead{$N_{\rm rej}$} & 
  \colhead{$\log_{10}({\rm sSFR/Gyr^{-1}})$}&
  \colhead{$N_{\rm bin}$} & \colhead{$N_{\rm rej}$} &
  \colhead{$\log_{10}({\rm sSFR/Gyr^{-1}})$}& 
  \colhead{$N_{\rm bin}$} & \colhead{$N_{\rm rej}$} &
  \colhead{$\log_{10}({\rm sSFR/Gyr^{-1}})$} 
  }
  \startdata
  \multicolumn{7}{c}{$M_{\rm stellar} = 5\times10^9\ M_\odot$}\\[0.3em]
  CSF no emission lines    & 
  54 & 2 & 0.50($\pm0.05$)     & 
  18 & 0 & 0.57($\pm0.08$)     & 
  73 & 3 & 0.42($\pm0.04$)     \\
  CSF with emission lines  & 
  51 & 2 & 0.55($\pm0.05$)     & 
  15 & 0 & 0.60($\pm0.06$)     & 
  48 & 5 & 0.54($\pm0.05$)     \\
  RSF no emission lines    & 
  67 & 0 & 0.49($\pm0.03$)     & 
  21 & 0 & 0.48($\pm0.04$)     & 
  78 & 4 & 0.60($\pm0.03$)     \\
  RSF with emission lines  & 
  68 & 0 & 0.51($\pm0.03$)     & 
  19 & 0 & 0.50($\pm0.04$)     & 
  71 & 5 & 0.65($\pm0.04$)     \\[.3em]
  \tableline\\[-0.7em]
  \multicolumn{7}{c}{$M_{\rm stellar} = 1\times10^9\ M_\odot$}\\[0.3em]
  CSF no emission lines    & 
  99 & 6 & 0.60($\pm0.04$)     & 
  32 & 4 & 0.66($\pm0.07$)     & 
  57 & 13 &0.79($\pm0.08$)     \\
  CSF with emission lines  & 
  100& 6 & 0.61($\pm0.04$)     & 
  35 & 4 & 0.68($\pm0.07$)     & 
  85 & 18& 0.80($\pm0.06$)     \\
  RSF no emission lines    & 
  117& 2 & 0.58($\pm0.03$)     & 
  36 & 2 & 0.67($\pm0.06$)     & 
  74 & 2 & 0.61($\pm0.04$)     \\
  RSF with emission lines  & 
  117& 2 & 0.61($\pm0.03$)     & 
  40 & 2 & 0.73($\pm0.07$)     & 
  81 & 3 & 0.73($\pm0.05$)
  \enddata
  \tablecomments{
    Mean values of the estimated ${\rm \log_{10}(sSFR/Gyr^{-1})}$ for
    our samples using different model assumptions. The sSFR was
    estimated for two stellar mass bins centered at $\log_{10}(M_{\rm
    stellar}/M_\odot)=9.7$ and 9.0. The width of each bin is $\pm0.3$ dex.
    The $N_{\rm bin}$ column indicates how many sources fall in each bin
    at each redshift for a given model.  Extreme values of the sSFR
    (${\rm \sim100~Gyr^{-1}}$, which correspond to minimum age models)
    were rejected before taking the mean. This latter choice does not
    make a significant difference (see Figure \ref{fig:SSFRcomparison}).
}
\end{deluxetable*}
}

\begin{document}
\title{Slow Evolution of the Specific Star Formation Rate at $z>2$:
The Impact of Dust, Emission Lines, and A Rising Star Formation
History }

\author{
Valentino Gonz\'alez\altaffilmark{1,2},
Rychard Bouwens\altaffilmark{3},
Garth Illingworth\altaffilmark{1},
Ivo Labb\'e\altaffilmark{3},
Pascal Oesch\altaffilmark{1,4},
Marijn Franx\altaffilmark{3}, 
\textsc{and}
Dan Magee\altaffilmark{1}
}
\altaffiltext{1}{Astronomy Department, University of California,
Santa Cruz, CA 95064}
\altaffiltext{2}{Department of Physics and Astronomy, University of
California, Riverside, CA 92521}
\altaffiltext{3}{Leiden Observatory, Leiden University , NL-2300 RA
Leiden, The Netherlands}
\altaffiltext{4}{Hubble Fellow}

\begin{abstract}

  We measure the evolution of the specific star formation rate (${\rm
  sSFR = SFR} / M_{\rm stellar}$) between redshift 4 and 6 to
  investigate the previous reports of ``constant'' sSFR at $z>2$, which
  has been in significant tension with theoretical expectations. We
  obtain photometry on a large sample of galaxies at $z\sim4\mbox{--}6$
  located in the GOODS South field that have high quality optical and IR
  imaging from \emph{HST} and \emph{Spitzer}. We have derived stellar
  masses and star formation rates (SFRs) through stellar population
  modeling of the rest-frame UV and optical spectral energy
  distributions (SEDs).  We estimate the dust extinction from the
  observed UV colors. In the SED fitting process we have studied the
  effects of assuming a star formation history (SFH) both with constant
  SFR and one where the SFR rises exponentially with time. The latter
  SFH is chosen to match the observed evolution of the UV luminosity
  function. We find that neither the mean SFRs nor the mean stellar
  masses change significantly when the rising SFR (RSF) model is assumed
  instead of the constant SFR model. We estimate the sSFR at $z>4$ for
  galaxies in two stellar mass bins centered at 1 and
  $5\times10^9\msun$. We find that the galaxies in the lower mass bin
  have very similar sSFRs to the more massive ones (within $\sim
  0.1\,$dex). When focusing on galaxies with $M_{\rm stellar}\sim5\times
  10^{9}\msun$, we find that the sSFR evolves weakly with redshift
  (${\rm sSFR(z) \propto(1+z)^{0.6\pm0.1}}$), consistent with previous
  results and with recent estimates of the sSFR at $z\sim2\mbox{--}3$
  using similar assumptions. We have also investigated the impact of
  optical emission lines on our results. We estimate that the
  contribution of emission lines to the rest-frame optical fluxes is
  only modest at $z\sim4$ and 5 but it could reach $\sim50$\% at
  $z\sim6$. When emission lines of this strength are taken into account,
  the sSFR shows somewhat higher values at high redshifts, according to
  the relation ${\rm sSFR(z)\propto (1+z)^{1.0\pm0.1}}$, i.e., the best
  fit evolution shows a value $\sim2.3\times$ higher at $z\sim6$ than at
  $z\sim2$.  However, the observed evolution is substantially weaker
  than that found at $z<2$ or that expected from current models (which
  corresponds to ${\rm sSFR(z)\propto (1+z)^{2.5}}$).

\end{abstract}


\section{Introduction}

Large samples of Lyman break galaxies (LBGs) have allowed for the study
of the properties of high redshift galaxies up to $z\sim8$ \citep[e.g.,
][]{bouw12,bouw11b, dunl12a, fink11, fink10, gonz10, gonz11, labb10,
  labb10a, lee12, mclu10, mclu11, oesc10, oesc12, papo11, scha10,
star09, wilk11}.  These samples are the result of large investments on
high quality \emph{Hubble Space Telescope} (\emph{HST}) data over deep
fields like GOODS \citep{giav04} that have rich complementary
multi-wavelength coverage.  Several studies have explored the observed
UV and optical colors of these galaxies. Through the standard technique
of synthetic stellar population modeling of the observed spectral energy
distributions (SEDs), the physical properties of these galaxies, such as
the stellar mass (\mstar), star formation rate (SFR), dust attenuation,
age, etc, have also been explored \citep[e.g., ][]{yan06, eyle05,
star09, gonz10, gonz11, papo11, lee12, bouw09, bouw12}.

These studies have shown that, at rest-frame UV and optical
wavelengths where LBGs are  more amenable to observations, the SEDs
show very similar colors with weak trends of bluer colors as a
function of decreasing UV luminosity and increasing redshift
(\citealt{bouw09, bouw12, wilk11, fink11}; but see also
\citealt{dunl12a}). As a direct consequence of this, the physical
properties estimated through SED fitting, using simple models, are
also remarkably similar.

Particularly intriguing are the results that indicate that the
specific star formation rate (sSFR) of sources with a given
\mstar~remains approximately constant from $z\sim2$ to $z\sim7$
\citep{star09, gonz10, mclu11}. In particular, \citet{gonz10} shows
that for sources with $M_{\rm stellar} = 5\times10^9\,\msun$, the sSFR
shows no evidence for significant evolution (${\rm sSFR }\sim 2\,
\gyr^{-1}$) from $z\sim7$ to $z\sim2$. Such a result is at odds with
the fairly generic theoretical expectation that the sSFR should
decrease monotonically with cosmic time \citep[e.g., ][]{wein11,
khoc11}.

Although suggestive, much of the early work pointing toward a
relatively constant sSFR was limited in many important aspects and
based on a number of simplifying assumptions. For example, early work
assumed zero dust extinction for the SFRs of $z>4$ galaxies. This was
motivated by the very blue UV colors observed in the galaxy SEDs,
which have subsequently been measured more accurately thanks to the
better wavelength coverage provided by \emph{HST}/WFC3 observations
\citep[e.g., ][]{bouw12}.  Early works also adopted exponentially
declining or constant star formation histories (SFHs). This is in
apparent contradiction with the evolution of the UV luminosity
function (LF), which suggests that rising SFHs are a better match for
the evolution of galaxies in early cosmic times \citep[e.g.,
][]{papo11}. Finally, much of the early work did not correct for the
effect of optical emission lines on the estimate of stellar masses
(\mstar), and hence on the sSFRs.

Several recent studies have attempted to redress some of these
shortcomings. \citet{bouw12}, for example, used new measures of the
dust extinction (based on UV colors) to correct the earlier sSFR
estimates at $z\sim4\mbox{--}7$. \citet{scha10} considered the
possible effects of the optical emission lines on the stellar masses
and sSFRs derived for high redshift galaxies, and this effort has been
considerably extended in \citet[see also \citealt{curt12}]{de-b12}.

In this paper we attempt to bring more consistency in the exploration
of the aforementioned issues, exploring the effect of dust reddening
in the UV colors, SFRs, and \mstar~measurements, as well as the
effects of choosing a SFH that better matches the evolution of the UV
LF, simultaneously. The goal is to make estimates for the physical
properties of $z>4$ LBG that use empirically motivated assumptions
that better match a larger range of observations. We plan to
investigate whether this new set of assumptions still shows an
approximate plateau in the sSFR at $z>2$ or shows evidence for an
increase toward high redshift as predicted by theory.

We now provide a brief plan for this paper. In Section 2, we briefly
describe the observational data and selection criteria used. In Section
3, we describe our approach to stellar population modeling, detailing
the specific assumptions that we make and the effects these assumptions
have on the physical properties we derive for LBGs. In Section 4 we
present the new measurements of the sSFR at high-redshift. We discuss
the results in 5, and summarize our findings in Section 6.

Throughout, we use a ($H_0,\,\Omega_M,\,\Omega_\Lambda$) = ($70\,\rm{km
  ~s^{-1}}$, 0.3, 0.7) cosmology when necessary and we quote all
magnitudes in the AB system \citep{oke83}.

\section{Data}

This work is based on a large sample of star forming galaxies in the
$z\sim4\mbox{--}7$ redshift range found using the Lyman Break technique
\citep{stei96} primarily in the ultra-deep HUDF/WFC3 field \citep[e.g.,
][]{oesc10}, and the deep Early Release Science (ERS; \citealt{wind11})
fields. For the $z\sim6$ ($i$-dropouts) search, we have also used the
more recent CANDELS GOODS-S data to obtain a larger sample. All these
fields have deep \emph{HST}/ACS \citep{giav04} and WFC3/IR imaging, as
well as the 23 h \emph{Spitzer}/IRAC data from the GOODS program
\citep{giav04}.  The sample has already been presented in
\citet{gonz11}, \citet{bouw07}, and \citet{bouw12} but we provide a
short summary here.

\subsection{\emph{HST} ACS and WFC3/IR photometry and sample
selection}

Over the ERS and the CANDELS GOODS-S field, both the ACS optical
($B_{435}$$V_{606}$$i_{775}$$z_{850}$) and the WFC3/IR
($Y_{098}$$J_{125}$$H_{160}$) images from \emph{HST} reach depths of
$\sim28$ mag ($5\sigma$ measured on 0\farcs35 diameter apertures). In
the HUDF field, the ACS optical data ($BViz$) are $1.5-2$ mags deeper,
and the WFC/IR data ($Y_{105}J_{125}H_{160}$) are 1.5 times deeper than
the ERS.

The \emph{HST} photometry for these sources was performed using the
SExtractor code \citep{bert96}. The images were registered to a common
frame and then PSF-matched to the $H_{160}$-band data. SExtractor was
run in double-image mode with the detection image constructed from all
images with coverage redward of the relevant Lyman break. Fluxes are
measured using Kron-style photometry. The criteria used to select
sources are as follows (for details and contamination rates please refer
to \citealt{bouw07}): 

$z\sim4~B$-dropouts:
$$(B_{435}-V_{606}>1.1)~\wedge~[B_{435}-V_{606}>(V_{606}-z_{850})+1.1]$$
$$\wedge~(V_{606}-z_{850}<1.6)$$

$z\sim5~V$-dropouts:
$$\{[V_{606}-i_{775}>0.9(i_{775}-z_{850})+1.9]~\vee~(V_{606}-i_{775}>2)\}$$
$$\wedge~(V_{606}-i_{775}>1.2)~\wedge~(i_{775}-z_{850}<1.3)$$

$z\sim6~i$-dropouts:
$$(i_{775}-z_{850}>1.3)~\wedge~(z_{850}-J_{125}<0.8).$$

The combined samples contain a total of 729 sources at $z\sim4$, 178 at
$z\sim5$, and 277 at $z\sim6$ (see table \ref{tbl:sampleSummary})

\tableSample

\subsection{\emph{Spitzer}/IRAC photometry}

While the \emph{HST}/ACS and WFC3/IR data allow us to find LBGs at
$z\gtrsim4$ and to study their rest-frame UV light, data from
\emph{Spitzer}/IRAC is needed to sample the rest-frame optical light
from such high redshift galaxies. These constraints are crucial to
derive reliable stellar masses from SED fitting \citep{papo01,
labb10}.

The HUDF and the ERS fields were imaged with \emph{Spitzer} in the four
IRAC channels as part of the \emph{Spitzer} GOODS program
\citep{dick03}.  In this work we only make use of the two most sensitive
channels centered at 3.6 and 4.5 \mum~respectively. The images have
integrations of $\sim23.3\,$h (the HUDF was imaged twice, thus amounting
to a total exposure of $\sim46.6\,$h). The depths of the images were
measured dropping 2\farcs5 diameter apertures in empty regions of the
image and determining the RMS. After applying a $1.8\times$ flux
correction to account for the light outside such aperture in the case of
a point source, the depths correspond to 27.8 mag ($1\sigma$) for a
single 23.3\,h image in the [3.6] channel and 27.2 mag for the [4.5]
channel.

The size of the IRAC PSF, however, is too broad to use standard
photometric techniques on our sample of faint LBGs. The flux within an
aperture is expected to be contaminated by flux from neighboring
sources spilling over onto the same aperture. The neighboring sources
are in general brighter and the contaminating fluxes can be
significant. The availability of higher resolution images (though at
different wavelengths) with \emph{HST} allows us to model the source
and its neighbors and subtract off the expected contaminating flux in
our apertures.  This method has already been described in several
previous works \citep{labb06, labb10, labb10a, gonz10, gonz11, wuyt07,
de-s07}. After the area around the source has been cleaned from the
flux of neighbors, we perform standard aperture photometry using
2\farcs5 diameter apertures.  An aperture correction is derived from
the higher resolution ``{\it template}'' image, in this case the
WFC3/IR $H_{160}$ image. This image is convolved to the resolution of
IRAC and an aperture correction factor is estimated for a 2\farcs5
diameter aperture. This correction factor is then applied to the
fluxes measured with IRAC and involves multiplying the flux of sources
(in 2\farcs5 diameter apertures) by 1.8 in both the [3.6] and [4.5]
IRAC channels, consistent with the correction one would apply to point
sources.

The cleaning procedure cannot always produce acceptable models of the
source and neighbors, resulting in poor subtractions for approximately
40\% of the cases. The individual \emph{cleaned} stamps were inspected
by hand and those with poor subtractions were excluded from the
subsequent analysis. This criterion mostly depends on the feasibility
of modeling the neighboring sources, and as a consequence, is not
expected to introduce any significant biases in our results.

\section{Stellar Population Modeling}

It has been customary to study the physical properties of high-redshift
galaxies by comparing their observed UV and optical SEDs with synthetic
stellar population (SSP) models. The quality of the flux constraints is
generally not adequate to constrain all model parameters simultaneously.
In particular, models with a variety of metallicities, initial mass
functions (IMF), and star formation histories can all reproduce the flux
constraints almost equally well. As a result, these model parameters
must be constrained independently or fixed to reasonable values.  In the
following analysis we have assumed a \citet{salp55} IMF with cutoffs at
0.1 and 100 $\msun$.

It is expected that the metallicities of high-redshift galaxies be
somewhat lower than that of local galaxies. For example, \citet{maio08},
from a small sample of $z\sim3.5$ galaxies determines gas metallicities
of $\sim0.2\,Z_\odot$.  A number of other observational studies have
also found similar metallicities \citep[e.g., ][]{pett00}, and
$0.2\,Z_\odot$ is also expected in many carefully-constructed
cosmological hydrodynamical simulations \citep[e.g., ][]{finl11}.  In
our analysis we therefore assume a fixed metallicity of $0.2\,Z_\odot$.
The resulting SFRs and stellar masses, which are the focus of this work,
are only weakly sensitive to this assumption.

For the SFHs, most early studies have generally assumed models with
SFRs that decline exponentially as a function of time with a
characteristic timescale $\tau$ left as a free parameter. The
particular case $\tau\rightarrow\infty$, which corresponds to models
with constant star formation (CSF), is sometimes used as the fiducial
model. Because in a smooth SFH the UV luminosity can be directly
linked to the SFR, these SFHs would predict a UV LF that shifts to
brighter luminosities at increasing redshift (or one that remains
constant in the CSF case), which is at odds with the current
determinations of the UV LF at $z>4$ \citep{bouw07, mclu09, oesc10,
bouw12}. If the SFHs, are in fact smooth, then the observed overall
dimming of the UV LF predicts a mean SFH that is better characterized
by a rising SFR \citep{papo11, star09, redd12}.  Oesch et al. (in
preparation), shows that the evolution of the UV LF can be well
reproduced by an exponentially rising SFH with a characteristic
timescale $\tau\sim500\,\myr$ \citep[see also ][]{smit12, papo11}. In
deriving the physical properties of galaxies in our samples, we will
look in detail at the extent to which these parameters depend on
whether we assume a constant or rising star formation (RSF) history.
We particularly examine the effects on the sSFR and its evolution with
redshift.

\subsection{Reddening}

\colormagB{figure}{ht!}{0.5}
\colormagRest{figure}{ht!}{0.5}

In addition to the necessary simplifying assumptions described at the
beginning of this section, other complications remain in the SED fitting
process. In particular, it is known that the effects of reddening
produced by dust and the reddening produced as a consequence of the
aging of the population are largely degenerate. This does not have a
substantial impact on the stellar masses which generally remain
constrained within $\sim0.3\,$dex for a range of ages, but the varying
dust corrections have a direct impact on the determination of the SFRs.

This degeneracy cannot be broken with the current data. However, one
potentially fruitful way forward for us is to not consider the full
range of stellar population models available to us (with ranging ages
and dust extinctions) and to estimate the dust extinction for individual
galaxies based on the UV continuum slope we observe using known
IRX-$\beta$ relations \citep[e.g., ][]{meur99}. Not only does this
approach have some justification based on observations of
$z\sim0\mbox{--}2$ galaxies \citep[e.g., ][]{meur99, redd12}, but
following this approach we are able to naturally reproduce the UV
color-luminosity and UV-to-optical color--luminosity relations.

Figures \ref{fig:colormagB}  and \ref{fig:colormagRest} show the
observed colors of our $z\sim4\mbox{--}6$ samples. The top panel of
Figure \ref{fig:colormagB} shows the observed $i_{775}-[3.6]$ color as a
function of observed [3.6] mag.  The different symbols are used to
divide the IRAC detected ($2\,\sigma$, solid blue circles), the marginal
detections ($1\mbox{--}2\,\sigma$, open squares), and the non-detections
(open gray circles). The histogram shows the color distribution of the
sample.  The two solid horizontal lines show the minimum color
(corresponding to a $>10\,\myr$ old population) and maximum color (set
by the age of the universe at $z\sim3.5$) of a CSF model in the absence
of dust reddening. The thick dashed line is a fit to the detected points
only and shows a trend to redder UV-to-optical colors for brighter
galaxies.

The middle panel of the figure shows the UV-slope characterized by
$\beta$ ($f_\lambda\propto\lambda^\beta$), as a function of the
magnitude in the [3.6] channel. A similar trend to redder colors for
brighter sources is observed. The UV-slope $\beta$ was measured by
fitting a power law to all the available \emph{HST}/ACS and WFC3/IR
photometric points for any given source (e.g., as done by
\citealt{bouw12} or \citealt{cast12}).

\citet{bouw12} study the dependence of $\beta$ with UV luminosity and
finds a similar trend \citep[see also ][]{fink11, mclu11}. As argued
in that work, $\beta$ can be affected by several factors but it is
most strongly dependent on the total dust content (which sets the
normalization applied to the dust curve assumed). 

The bottom panel of Figure \ref{fig:colormagB} shows the result of
applying a dust reddening correction derived solely from the UV-slope to
the UV-to-optical color. We find that the uncorrected trend in
UV-to-optical colors versus [3.6] is best fit by the relation
$(i_{775}-[3.6])=-0.25 \times [3.6]+{\textit const}$. After applying the
dust correction the trend is now flatter $(i_{775}-[3.6])_{\rm
dust-corrected}=-0.10 \times [3.6]+{\textit const}$ and most sources in
the sample are in the color range that is covered by our dust-free CSF
models \citep[see also ][]{oesc12a}. The trend in UV-to-optical colors
(after removing the effect of dust) still shows a slight dependence on
the [3.6] luminosity, but this is probably due to the fact that sources
that are brighter in the rest-frame optical are on average older. This
is similar to the results shown in \citet{labb07}, who show that the UV
and UV-to-optical color trends exhibited by star forming galaxies in the
$z\sim 0.7\mbox{--}3.5$ range can be primarily explained by the effects
of dust reddening, with a small contribution due to age.  Figure
\ref{fig:colormagRest} shows a very similar behavior at $z\sim5$ and 6.

In view of this result, and to alleviate the difficulties produced by
the age-dust degeneracies, in the following modeling we establish the
dust reddening directly from the observed UV slope $\beta$ following
the \citet{meur99} relation:
\begin{equation}
  \label{eq:meurer}
  A_{1600} = 4.43 + 1.99 \beta
\end{equation}
where $A_{1600}$ is the attenuation in magnitudes at 1600 \AA. The
attenuations at other wavelengths are determined using a
\citet{calz00} extinction curve. Sources with $\beta<-2.23$ are
assigned $A_{1600}=0$.

\subsection{Photometric redshifts}

Spectroscopic redshifts at $z>4$ are sparse and generally biased towards
bright sources or sources with Ly$\alpha$ emission. Given the limited
spectroscopic sample and their current biases we will instead rely on
photometric redshifts derived from our SED fitting with the code FAST
(Kriek et al. 2009).

The models used to estimate these redshifts correspond to the
\citet{bruz03} (BC03) stellar population models with constant star
formation history (described in more detail below). These models do not
incorporate emission lines.

One of the goals of the current study is to assess the effect that
emission lines can have on the properties derived through SED fitting,
in particular on the sSFR. Adding emission lines to the models can have
important effects on the photometric redshifts that will also impact the
stellar masses. For example, if emission lines are included in the
models, galaxies with significantly blue observed [3.6]-[4.5] colors are
more likely to be placed at a redshift where emission lines can
contribute to the IRAC photometry. This is because the optical colors of
the stellar continuum are usually red and only the addition of emission
lines can produce blue colors -- at particular redshifts. If the
photo-$z$ places the galaxy at a redshift were the IRAC photometry is
contaminated by emission lines, the derived stellar masses will be lower
compared to the case without emission lines.

A problem arises because the strengths of the lines and the line ratios
assumed and added to the models have a directly impact on the
photometric redshift likelihood and, as a consequence, on the stellar
masses. We would like to study the effects of different assumptions
for the strengths of the emission lines and it would be useful to
disentangle the effects of the emission lines on the photometric
redshift and on the stellar masses.

Since we will concentrate mostly on the effects that emission lines have
on stellar mass and SFR, we have chosen to derive photometric redshifts
(and confidence intervals) excluding the IRAC photometry, effectively
removing the possible effect of optical emission lines on the
photometric redshifts. Hence these redshifts are mostly driven by the
observed wavelength of the Lyman break. The photo-$z$s are only derived
once and are used throughout (in particular, the same photo-$z$ are used
for all the modeling variations explored in the paper).

Figure \ref{fig:zdist} shows the redshift distributions obtained for
each sample (we have ignored secondary redshift solutions in the
photo-$z$). These are in very good agreement with the expectations from
the LBG selection \citep[e.g., ][]{bouw07}.

\zdist{figure}{ht!}{0.5}

\subsection{The Constant Star Formation model}

\sfrvsmstar{figure}{ht!}{0.5}
\SFRvsSFRnew{figure*}{ht}{0.9}

We have used the FAST \citep{krie09} SED fitting code to fit the
observed rest-frame UV + optical SEDs of the galaxies in our sample with
a set of synthetic stellar population models (we use the
\citealt{bruz03} models). In this section we present the results
obtained when a smooth constant star formation (CSF) rate history is
assumed. We have assumed a metallicity $Z=0.2\,Z_\odot$, and a
\citet{salp55} IMF with cutoffs at 0.1 $\msun$ and 100 $\msun$. The
photometric redshifts (and the confidence intervals) used throughout the
modeling do not include the IRAC photometric measurements or constraints
(see Section 3.2). The set of models considered in this section only
include fits to the stellar continuum fluxes. They do not include
emission lines. Models that include emission lines are considered later.

Additionally, and as described in previous sections, we do not leave the
reddening by dust as a free parameter in our modeling, but rather fix it
based on the observed UV slope $\beta$ (Equation \ref{eq:meurer}). In
doing so, we greatly simplify  the stellar population modeling of
high-redshift galaxies by eliminating the very significant degeneracy
between dust and age. While the impact of this degeneracy on the derived
stellar mass is more limited, this degeneracy can have a big impact on
the SFRs. For example, at $z\sim4$, a model with zero dust reddening can
produce SFRs that are lower by 1.8 dex compared to a model with maximal
dust reddening (minimum age). Meanwhile, the effect on the stellar mass
is generally $\lesssim 0.3$ dex.

The results of the SED fitting procedure are shown in Figure
\ref{fig:sfrvsmstar}. The figure shows the SFR versus \mstar~for the
$z\sim4$ (top, blue), $z\sim5$ (middle, green), and $z\sim6$ (red)
sources. The dashed lines show the minimum and maximum values of
\mstar~at a given SFR that our models allow for. The distributions are
very similar at all redshifts, suggesting little evolution of this
relation, consistent with previous results that use similar
assumptions \citep{star09, gonz10, gonz11}.  It should be noted,
however, that in previous works, the relation presented generally
corresponds to SFRs that have not been corrected for dust extinction.
Figure \ref{fig:sfrvsmstar} shows the intrinsic SFRs, i.e., after
correction for dust reddening and dimming. The stellar masses have
been derived from the same models used to derive the SFR. The fact
that the relations still show little evolution is a consequence of the
UV and optical colors being similar, implying similar properties of
the galaxies as a function of redshift, including dust corrections
\citep[see also ][]{gonz12, bouw12}.

A small fraction of the sample shows ages that are extremely young,
crowding the $10\,\myr$ dashed line. These sources are generally sources
that are marginally detected or non-detected in the IRAC channels. The
very blue UV-slopes of these sources dominate the fitting, driving them
to young ages. It is possible that they are indeed very young but this
is hard to establish with the depths of the IRAC data used here (see
also \citealt{oesc12a}, where the fraction of young sources at $z\sim4$
is estimated to be very small at the bright end based on the rest frame
UV-to-optical colors). The analysis in this paper focuses on the more
massive sources that are detected in $Spitzer$/IRAC, so this small
fraction of the population should not bias our main results.

In the following sections we explore the impact of two important
ingredients that have not been taken into account in our previous
modeling. We use the results from the CSF model as our base for
comparison. Our first major consideration is the impact of assuming a
rising SFH (which reproduces better the evolution of the UV LF at $z>3$)
on our SFR and stellar mass estimates. Our second major consideration is
to explore the effects of emission lines on the derived physical
parameters.

\subsection{The Rising Star Formation Model}

\mstarmstar{figure*}{ht!}{0.9}
\emlineCorrections{figure*}{ht!}{0.9}

Under the assumption of a smooth star formation history, the UV
luminosity of galaxies is related to their SFR (e.g., the Madau
relation; \citealt{mada98}; see also the SFR functions derived from the
UV luminosity functions -- \citealt{smit12}). The exact form of the
relation does not change strongly between different smooth SFHs if the
ages considered are older than $\sim100\,\myr$.  Based on this relation,
the previously presented model with a CSF history, makes a clear
prediction for the evolution of the UV LF: it predicts a LF that does
not change with redshift.  This has been thoroughly ruled out using
large samples of LBGs at $z>4$ \citep[e.g., ][]{bouw07, bouw11b,
oesc10}.

The observed UV LF evolves with cosmic time showing a brighter
characteristic magnitude at lower redshifts at least until $z\sim3$.
A model that better matches the observed evolution is one in which the
average SFR of galaxies rises with time \citep[e.g, ][]{papo11,
redd12, smit12}, something that has also been predicted in several
numerical simulations \citep[e.g][]{finl11, jaac12a}. 

By following sources at a constant cumulative number density
$n(<M_{UV}) = 2\times 10^{-4} {\rm Mpc^{-3}}$ as a function of
redshift, \citet{papo11} derive a best-fit exponentially rising SFH of
the form ${\rm SFR \propto e^{(t-t_0)/\tau}}$, with $\tau=420\,\myr$
(they find a slightly better fit using a linear model but the
differences are not large). A similar analysis at multiple values for
the number density yields a similar result, with a best fit
$\tau\sim500\,\myr$ at all densities (Oesch et al. in preparation).
The goal of the following analysis is to study the effects of such a
model on the determination of the stellar masses and star formation
rates. The details of the star formation history used are not
important as long as it agrees (or at least agrees better than the CSF
model) with the evolution of the UV LF.

We have re-determined the SFRs and stellar masses for the galaxies in
our sample using a RSF model as described before (${\rm SFR \propto
e^{(t-t_0)/\tau}}$) , with $\tau=500\,\myr$. The parameter $t_0$ is
free in our model. For the RSF and the CSF models, the SFR can be
derived directly from the UV luminosities (if the dust extinction is
known -- here we assume it can be derived from the UV-slope). In fact,
Figure \ref{fig:sfrvssfr} shows that, as expected, the SFRs derived
from the CSF and the RSF model are on average the same with little
scatter.  Since the final goal is to study the sSFR in our sample, we
now focus on the effects of the RSF on the stellar masses.

As can be seen in the Figure \ref{fig:mstarmstar} (top row) there is a
significant scatter ($\sim0.25\,$dex) in the comparison between the
CSF and the RSF stellar masses. The mean value, however, does not
change significantly when the IRAC detected sources are considered
(solid circles). The IRAC undetected sources, meanwhile, show a slight
bias towards larger RSF masses, but this could be attributed to the
poor constraints in the rest-frame optical, which is very important to
derive stellar masses. This is consistent with a recent report by
\citet{redd12}, who also find that the average \mstar~does not change
between models with CSF and RSF for a sample of galaxies at
$z\sim2\mbox{--}3$.  It is however, inconsistent with the reports of
\citet{papo11}, who finds larger stellar masses when a RSF is assumed.
In their analysis, however, the age is fixed with a formation redshift
$z=11$, artificially fixing the ages of the sources.

There is one important observation regarding the RSF models. A
significant fraction of the sources that we have modeled result with
ages that correspond to the maximal allowed in our set of models. This
is a result of the fairly red UV-to-optical colors exhibited by these
galaxies (after dust reddening corrections). The models considered
here only include stellar continuum light, and as a consequence, these
red colors can only be interpreted as older ages.  Two possible ways
to alleviate this tension are: the presence of emission lines that
would contribute to the flux measured with $Spitzer$/IRAC
\citep{scha10}; or a non-smooth component to the SFH such that the SFH
is rising on average but there is a characteristic duty cycle that
make the colors redder at any given age (\citealt{labb10}; see also
recent simulation results by,  e.g., \citealt{jaac12}).  Since many of
the sources that present maximal age at $z\sim4$ are expected to be
free of emission line contamination given their redshifts, the actual
solution to the problem is probably a combination of both scenarios.

In summary, both the SFRs and the stellar masses obtained from the RSF
are on average the same as for the CSF model. We would therefore not
expect the average sSFR to be substantially affected by whether one
adopts a RSF or CSF history for the stellar population modeling.

\subsection{The possible impact of optical emission lines.}

We turn now to the effect of the possible contribution of emission
lines to the rest-frame optical photometry.  These lines are not
expected to affect significantly the SFRs derived from SED fitting but
they can, in principle, have a strong effect on the values derived for
the stellar masses.

It has been suggested that a (possibly large) fraction of the flux
detected with $Spitzer$/IRAC from galaxies at $z>4.5$ could be coming
from nebular regions associated with star formation in the form of
emission lines \citep[e.g., ][]{scha10}. Unfortunately, it is not
possible to directly observe these emission lines in $z>4$ galaxies
with spectroscopy using current facilities.

Nonetheless, we can infer the approximate strength of these emission
lines indirectly. For example, in a recent study, \citet{shim11} used a
spectroscopic sample at $3.8<z<5.0$ to study the H$\alpha$ EW from broad
band photometry. In this redshift range, H$\alpha$ falls in the IRAC
[3.6] filter, adding to the rest-frame optical continuum flux at this
wavelength. However, H$\alpha$ does not contribute to light in the
adjacent K-band or IRAC [4.5] filters, making it possible to infer the
EW of this emission line. \citet{shim11} find a median H$\alpha$ EW of
$\sim480\,$\AA.  This technique, however, is only sensitive to objects
with large H$\alpha$ EW ($>350\,$\AA). While this experiment shows that
there is a fraction of the star-forming population at $z\sim4$ with
strong line emission, the objects selected by \citet{shim11} are not
expected to be typical for the full population.  In a recent similar
study, \citet{star13} find a lower mean H$\alpha$ EW = 270 \AA~in the
same redshift interval.  Independent evidence for extreme line emitters
at $z\gtrsim2$ come from \citet{atek11} who identify a large sample of
such objects at $z\sim2.5$ based on \emph{HST} WFC3/IR Grism data.

Since the direct observation of emission lines at these redshifts is not
currently possible, we need to make assumptions based on our
understanding of these lines at lower redshift. At $z\sim2-2.5$, for
example, \citet{erb06} find that the rest-frame EW of H$\alpha$ for
galaxies with $\log_{10}(M_{\rm stellar})\sim10.0\mbox{--}10.5$ is
$\sim100\,$\AA.  The EW shows an increase with decreasing \mstar.
Moreover, this EW also seems to increase toward higher redshift. In a
study based on \emph{HST} Grism spectroscopy from the 3D\emph{HST}
survey \citep{bram12}, \citet{fuma12} find that the H$\alpha$ EWs of
$10^{10}\mbox{--}10^{10.5}~{\rm M_\odot}$ galaxies at
$z\sim0\mbox{--}2.5$ are best fit by the following relation:
\begin{equation} \label{eq:mattia} EW(z) \sim 15.8\times (1+z)^{1.52}
\rm{\AA}, \end{equation}

The typical strength of the optical emission lines at higher redshifts
and lower masses will likely remain uncertain for some time. In the
following, we estimate the effects that seem plausible based on these
observations.

We take two different approaches. In our first, more conservative
approach, we will assume that the mean H$\alpha$ EW observed at
$z\sim2\mbox{--}2.5$ \citep{erb06} remains constant at 100\,\AA~at all
redshifts $z>3$ and stellar masses. Given the trends in the EW described
earlier with stellar mass and redshift, this is likely an underestimate
of the EW exhibited by real galaxies at very high-redshift. A more
extreme approach is to assume that the trends observed at $z\sim0-2.5$
continue to higher redshifts following the same extrapolation ( Equation
\ref{eq:mattia}).  We call this the \emph{maximal} emission line model.
For each of the models (\emph{conservative} and \emph{maximal}) we
derive the strength of all the other emission lines based on the flux
ratios from \citet{ande03} assuming a $Z=0.2\,Z_\odot$ metallicity and a
flat $F_\nu$ underlying optical continuum (i.e. flux ratios between the
lines correspond exactly with EW ratios). Then we calculate the
contribution of the emission lines to the broadband magnitudes assuming
the redshifts derived from the CSF models in Section 3.2. Figure
\ref{fig:emlineCorrections} shows the emission line contributions (in
magnitudes) that we estimate as a function of redshift. The solid
histograms (arbitrary normalization) show the redshift distribution of
our samples.

In a previous study \citep{gonz11}, we examined the median SEDs of this
sample by stacking the photometry in bins of UV luminosity.  The SEDs of
$z>4$ galaxies show a consistent excess in their [3.6] flux over their
[4.5] flux which, as argued in that work, can be explained by the effect
of emission lines. It was shown in that work that a simple model with
H$\alpha$ EW = constant $=300\,$\AA~can simultaneously reproduce the
observed UV-to-optical and $[3.6]-[4.5]$ colors exhibited for the sample
(the corresponding EWs for the other most prominent optical lines are
${\rm EW_{rest}(}$\ion{O}{2}, H$\beta$, \ion{O}{3})=(189, 105,
670)\,\AA, respectively). This EW is very similar to the value predicted
by the \emph{maximal} model at $z\sim6$.

Next, we subtract the emission line contribution from the photometry,
and we refit the SEDs with CSF and RSF models. As expected, the SFRs
derived in this way are unchanged in both cases (Figure
\ref{fig:sfrvssfr}, bottom row). This is because the SFRs of star
forming galaxies (with ages $\gtrsim100\,\myr$) depend almost
exclusively on the UV fluxes, which are unaffected by these emission
lines. The effects on the stellar masses can be seen in Figure
\ref{fig:mstarmstar} (bottom row). This figure only shows the effects
for the \emph{maximal} model.  The effects from the
\emph{conservative} model are much smaller.

The impact of the emission lines on the stellar mass of galaxies is
largest for the $z\sim6$ sample. This is expected, since at this
redshift there are strong emission lines affecting both the [3.6]
(H$\beta$ and \ion{O}{3}) and the [4.5] (affected by H$\alpha$)
channels. The average change in stellar mass estimates $\Delta{M_{\rm
stellar}}$ for IRAC detected sources at this redshift is $\sim0.26\,$dex
for the \emph{maximal} model (and only 0.1 dex for the
\emph{conservative} model).  Correcting for emission lines results in
lower stellar masses and higher sSFRs for galaxies \citep[see also
][]{curt12}. The impact of this on the evolution of the sSFR, however,
is not straightforward to assess, since correcting for the effect of
the emission lines on the mass shifts sources to lower mass where the
sSFR also may be higher.

\highzSSFR{figure*}{ht!}{0.80}
\tableSSFR{}

\section{The sSFR at $z\gtrsim2$}

In the preceding sections we have studied the effects that different
stellar population modeling assumptions have on the SFR and
\mstar~estimates for high-redshift LBGs. We have shown that the SFRs
do not depend on whether a rising SFH or CSF is considered in the
modeling. The SFRs are also not significantly affected when emission
lines, which affect the rest-frame optical fluxes, are taken into
consideration.  For the simple smooth RSF that we consider, the
average \mstar~derived for these galaxies does not change
systematically with respect to the CSF model (although there is a
scatter of $\sim0.25\,$dex in the values determined). When the
emission lines are considered, however, the stellar masses could be
$\sim0.26$~dex smaller at $z\sim6$. 

Figure \ref{fig:SSFRcomparison} shows the sSFR determined from our
sample as a function of redshift $z\gtrsim4$ (see also Table
\ref{tbl:SSFRresults}).  The left panels shows the mean sSFRs derived
from a CSF model and the right panels show the results using the
exponentially rising SFH. The top panels correspond to stellar
continuum only models and the bottom panels consider the effects of
emission lines assuming the average contribution expected from the
\emph{maximal} model. The mean sSFRs were determined for galaxies in
two different \mstar~bins. The blue circles and green hexagons
correspond to galaxies with $M_{\rm stellar,CSF} \sim 5\times 10^9\msun$
(the width of the bins are $\Delta\log({M_{\rm stellar}}) =
\pm0.3\,$dex).  The orange pentagons and purple triangles show the
sSFR for galaxies that are $\sim5\times$ less massive. 

Galaxies with best fit ages $\lesssim10$ Myr (the minimum in our grid of
models) were rejected before estimating the mean values. The
corresponding sSFR for the galaxies that were rejected is ${\rm sSFR
\sim 100~Gyr^{-1}}$ (Table \ref{tbl:SSFRresults} indicates the number of
sources rejected in each bin). The effect of keeping such galaxies is
very small as illustrated for the more massive bin by the gray symbols
in Figure \ref{fig:SSFRcomparison}.

Comparing the mean sSFRs estimated for both mass bins, the data may
suggest that the less massive galaxies present slightly larger sSFRs.
However, the differences are small ($\lesssim0.1\,$dex) and very
uncertain.  For both mass bins and regardless of the modeling
assumptions, the sSFR shows very little variation with redshift from
$z\sim4$ to $z\sim5$, consistent with previous reports \citep{star09,
gonz10, mclu11}. However, there are important differences among the
$z\sim6$ determinations. These variations mostly reflect the fact that
the sSFR at $z\sim6$ is highly sensitive to the modeling assumptions.

It should be noted that the galaxies that make up the $1\times10^9\
M_\odot$ and $5\times10^9\ M_\odot$ samples are not the same galaxies
in the top and bottom panels. In the top panel the \mstar~from CSF
models with no emission lines is considered whereas in the bottom, the
masses used correspond to the ones derived when the \emph{maximal}
model of emission lines is assumed. This is important because if the
same galaxies were considered, then their sSFRs should be larger when
the emission lines are considered (because their masses are lower).
Most importantly, this effect would be stronger at $z\sim6$,
suggesting a stronger evolution of the sSFR with redshift. This does
not seem to be the case when galaxies within the same mass bin are
compared.

The mean sSFR values derived with all the model assumptions are
approximately $\sim3\,\gyr^{-1}$, slightly larger than previous reports
(without dust corrections or emission lines) and the values reported at
$z\sim2\mbox{--}3$ \citep[see Figure \ref{fig:sSFR}]{dadd07, dadd09,
redd12}. A weighted best fit to the evolution of the sSFR as a function
of redshift is: $\log({\rm sSFR}(z)/{\rm Gyr^{-1}}) = -0.1(\pm0.1) +
1.0(\pm0.1) \log(1+z)$. This fit was obtained considering the low
redshift results from \citet[$z\sim2$]{dadd07} and
\citet[$z\sim2\mbox{--}3$]{redd12} as well as the values derived here
for the mean sSFR at $z\gtrsim4$ (taking the RSF with \emph{maximal}
emission line model as our fiducial values). The weights used correspond
to the random errors.  Since the random errors are not available for
many determinations in the literature, the weights were normalized based
on the sizes of the samples. Our result suggests a mean sSFR that
increases slightly with redshift, however, the evolution observed is
much weaker than that seen at $z<2$. It is also inconsistent with the
slope expected from theoretical models which generally follow closely
the specific halo accretion rate, resulting in ${\rm sSFR}(z) \propto
(1+z)^{2.5}$ \citep[e.g., ][]{wein11, bouc10, dave08}. If the CSF model
without emission lines is assumed instead, the derived evolution is
flatter: ${\rm sSFR}(z) \propto (1+z)^{0.6\pm0.1}$

\section{Discussion}

Many of the earliest studies of the sSFR at high-redshift had suggested
that the sSFR did not evolve strongly at $z\gtrsim2$ \citep{gonz10,
star09, mclu11}. This is very different from the behavior observed at
lower redshifts and also from the expectations of theoretical models
that consistently predict a sSFR that declines monotonically with cosmic
time when halos of a constant mass are studied \citep[e.g., ][]{wein11,
bouc10, dave08}. Most earlier studies, however, did not determine the
SFRs and \mstar~in a way that was clearly self-consistent, did not adopt
SFHs which are consistent with the evolution of the UV LF, and also did
not account for the effects of nebular emission lines in the stellar
population modeling.

More recently, however, there have been some efforts to redress these
shortcomings. For example, \citet{bouw12} focus on correcting previous
sSFR determinations to reflect the latest estimates of dust extinction.
Another example is the work of \citet{de-b12} where there is an effort
to correct for the impact of the emission lines on the inferred stellar
masses. \citet{star13} also explores the effect of emission lines using
a modeling technique very similar to the one presented here. In general,
all these recent studies, including the present work, have suggested
somewhat higher sSFRs at $z\gtrsim4$. It is worth noting, though, that
despite the growing consensus that the SFRs at $z\gtrsim4$ increase,
there is fairly large difference in the magnitude of evolution derived.
In the following we compare to a couple of recent results with the goal
of highlighting the main analysis differences that yield to the
different results.

\subsection{Comparison to \citet{de-b12}}

In \citet{de-b12} the authors find an order of magnitude larger sSFRs at
$z>4$ than is found at $z\sim2$, the sSFRs we find at $z\gtrsim4$ are
only larger than the $z\sim2$ values by factors of
$\sim1.5\mbox{--}2.0$. What are the reasons for these differences?
Largely, this is the result of the different assumptions used in
modeling the observed photometry. Differences in assumptions can have an
impact on the derived sSFRs due to degeneracies between different model
parameters.  While we adopt the simplifying assumption that the observed
UV continuum slope can be used to estimate the dust extinction in
high-redshift galaxies, \citet{de-b12} do not impose any constraints on
the dust, metallicity, mass, or age in modeling the photometry of
high-redshift galaxies. By not restricting the model parameter space
through various simplifying assumptions, \citet{de-b12} observe
considerable scatter in the properties of many of their sources and also
find a large fraction of sources to have very young ages (a result which
we consider probably unphysical and may result in \citealt{de-b12}
reporting very high sSFRs). On the other hand, with our approach, we use
the UV continuum slope to set the dust extinction, effectively breaking
many of the model degeneracies.  While one can debate which approach
yields the most accurate results, we prefer our approach due to the
strong evidence at both $z\sim0$ and $z\sim2$ that the UV continuum
slope correlates on average with the observed dust extinction
\citep{meur99, burg05, over11, dadd07, redd12}.

\sSFR{figure*}{ht!}{0.95}

\subsection{Comparison to \citet{star13}}

\muvmstar{figure*}{t}{0.85}

In a recent study \citet{star13} explores the impact of rest-frame
optical emission lines on the stellar masses and sSFR derived through
SED fitting at $z\gtrsim4$. Their analysis is similar in many regards to
the one presented here but they find a sSFR evolution that is much
faster with redshift, in agreement with theoretical expectations.  As we
discuss below, it appears that the reason for this difference is that
the consideration of the effect of the $M/L$ scatter turns out to play
an important role.

The SED modeling assumptions used in \citet{star13} are very
similar to the ones we have used here, and the resulting SFR and
stellar mass estimates for individual sources are in good agreement,
as can be seen in Figure \ref{fig:muvmstar}. Only sources that are
detected in IRAC ($>2\,\sigma$) are shown in the figure (background
points). The large open squares correspond to our estimates for the
median SEDs stacked by UV luminosity from \citet{gonz11} and the
dashed lines correspond to the trends derived by \citet{star13} based
on the same stacked SEDs.  This figure makes it clear that there is
broad overall agreement in the basic properties derived.

The way in which the average sSFR is determined at a given redshift,
however, causes important differences in the conclusions. The results
presented in the \citet{star13} work are based on the best fit
$\log(M_{\rm stellar})\mbox{--}M_{\rm UV}$ relation derived from the
stacked SEDs. These stacked SEDs are binned according to their UV
luminosity. Their average sSFRs at a given redshift, then, correspond to
UV luminosity binned averages. This is similar to the result by
\citet{bouw12}, who only apply an improved dust correction to the
luminosity binned results but \citet{star13} also include mass
corrections due to the effect of emission lines.

In the present analysis we estimate the sSFR in bins of stellar mass.
Within a given bin there will be galaxies with lower than average
luminosities but high $M/L$ ratios, as well as bright galaxies with low
$M/L$ ratios.  Since fainter galaxies are more numerous, the
distribution of $M/L_{\rm UV}$ ratios within the bin is skewed to high
$M/L$ ratios (equivalent to lower sSFRs, see large filled squares in
Figure \ref{fig:muvmstar}).

\citet{star13} recognize and discuss the differences that the $M/L$
ratio distribution at a given mass cause on the mean sSFR \citep[see
also][]{redd12}. They report that, at $z\sim4$, the mass binned average
sSFR could be $2.8\times$ lower if a symmetric scatter of 0.5\,dex (the
observed scatter reported in \citealt{gonz12}) was assumed instead of no
scatter. Nevertheless, they report their results assuming zero
  scatter in $M/L$ and suggest that the above effect could be offset by
  possibly higher UV luminosity to SFR conversion factors at higher
  redshift. Meanwhile, in our estimates we consider the \emph{observed}
  $M/L$ distribution at a given mass. Neither approach is perfectly
  correct, but this explains why our estimates of the sSFRs are lower.

Ideally, we should use the intrinsic distribution of $M/L$ ratios,
  as opposed to the observed distribution which is broader due to
  modeling and observational uncertainties. A reliable estimate of the
  intrinsic scatter in M/L ratios is out of the scope of this paper
  given the absence of rest-frame optical spectroscopy, and will likely
  have to wait until JWST. Here we just note that our results reported
  in Table \ref{tbl:SSFRresults} are consistent with those of
\citet{star13} if a symmetric intrinsic scatter of 0.3\,dex in
$\log(M_{\rm stellar})$ is added to their mean relation at $z\sim4$ (0.2
dex at $z\sim6$).

It is interesting to note that in the works of \citet{de-b12},
\citet{star13}, as well as our study, the sSFR evolution from $z\sim4$
to $z\sim5$ is quite small (for a given set of assumptions, in
particular, for SFH and emission line strengths). As shown in Figure
\ref{fig:mstarmstar}, at these redshifts emission lines have only a weak
effect on the stellar masses and SFRs.  Differences in the sSFR appear
only at $z>5$, which is also when emission lines can have a bigger
impact on the stellar population modeling. Furthermore, there is only
$\sim240$ Myr between $z\sim5$ and $z\sim6$. Even though it is possible
that the differences in sSFR arising between $z\sim6$ and $z\sim5$ are
real, it seems surprising that significant changes manifest themselves
at the same redshift where emission lines potentially play a large role
in the SED modeling (especially considering how uncertain the assumed
EWs are).

We conclude that the sSFR \emph{at a constant stellar mass} changes only
weakly with cosmic time when full consideration is given to the scatter
in $M/L$ -- with or without the consideration of emission lines (Figures
\ref{fig:SSFRcomparison} and \ref{fig:sSFR}). A fit to the
\emph{maximal} emission line estimates of the sSFR, which includes the
values at $z\sim2\mbox{--}3$ reported in the literature \citep{dadd07,
redd12}, yields a best-fit to the evolution of the ${\rm sSFR} \propto
(1+z)^{1.0\pm0.1}$, indicating that the best fit sSFR is higher by
$\sim2.3\times$ at $z\sim6$ than at $z\sim2$ (compared to the factor
$\sim8.3\times$ predicted in most simulations). When emission lines are
not included in the models (as in our CSF model), the exponent is lower:
0.6$\pm0.1$. The evolution we derive at $z\gtrsim2$ is less than
expected from the lower-$z$ trends and is not consistent with the
predictions from numerical simulations \citep[e.g., ][]{neis08, wein11}.

\section{Summary}

We take advantage of the ultra-deep and wide-area ACS + WFC3/IR +
\emph{Spitzer} observations of the GOODS-S field to derive flux
measurements for a sizable sample of $z\sim4\mbox{--}6$ galaxies and use
these measurements to more thoroughly quantify their SFRs, stellar
masses, and other properties. We have explored the effects of using a
reddening law based on the UV colors only \citep{meur99} and
investigated the impact of the star formation history by alternatively
considering constant star formation (CSF) and smoothly rising star
formation (RSF) models. We have also studied the effects of optical
emission lines on the SFR and stellar masses assuming two different
models for the emission line strengths. Our goal has been to quantify
the impact of these assumptions in estimating the specific star
formation rate of LBGs which has previously been reported to show little
evolution at $z\sim2\mbox{--}7$ \citep{gonz10, star09, mclu11}.  The RSF
history explored here matches the evolution of the  UV LF \citep[e.g.,
][]{papo11, smit12} and is in good agreement with the predictions from
recent smooth particle hydrodynamics simulations \citep{finl11}. Our
main findings are the following:

\begin{itemize}

  \item At a given redshift, LBGs that are fainter in the rest-frame
    optical show bluer UV colors. This is consistent with the previously
    reported trends of bluer UV colors for fainter galaxies
    \citep{bouw09, bouw12, wilk11}. The UV-to-optical colors are also
    bluer for fainter sources \citep{papo04, gonz12}. Such
    luminosity-dependent trends could arise from a dependence of the
    dust content on mass (or luminosity) or from redder stellar
    populations as a result of aging with a smooth SFH \citep[see
    also][]{labb07, bouw09, bouw12, gonz12}.

  \item Assuming that the UV slope reddening is caused by dust alone
    as in Equation \ref{eq:meurer} (\citealt{meur99}; and following a
    \citealt{calz00}, dust law), the dust-corrected UV and
    UV-to-optical colors can be reproduced by dust-free CSF models. A
    residual trend towards bluer colors at lower luminosities can be
    caused by the effects of aging of the population \citep[see
    also][]{gonz12}.

  \item We find that the SFRs derived assuming a CSF history do not
    change when a smoothly rising SFH (RSF) is used instead. The RSF
    that we use is chosen to match the evolution of the UV LF (${\rm
    SFR \propto e^{(t-t_0)/\tau}}$; $\tau=500\,\myr$). The mean
    stellar masses are also unchanged when the RSF is assumed,
    although there is a significant scatter of $\sim0.25\,$dex in the
    ${\rm M_{stellar,CSF} - M_{stellar, RSF}}$ relation.

  \item We explore the effects of emission lines assuming two
    different models for the strength of the lines as a function of
    redshift. Regarding the SFRs and stellar masses, the impact of
    adding emission lines with strengths consistent with the trends at
    lower redshifts is important, but modest. \emph{The inclusion of
    emission lines only affects our derived stellar masses}. For the
    \emph{maximal} emission lines model, the stellar masses are up to
    0.26\,dex lower at $z\sim6$ (and the sSFR larger; see also
    \citealt{curt12}). For the model with lower emission line
    strengths, the masses change by only $\sim0.1$ dex.

  \item We estimate the sSFR for galaxies in two stellar mass bins in
    our samples at $z\sim4, 5$, and 6. There is a hint that the sSFRs of
    lower mass sources may be slightly larger than those of more massive
    systems. In particular, for the CSF model, $1\times10^9\ M_\odot$
    galaxies have sSFRs that are $\sim0.1\,$dex larger than sources with
    $5\times10^9\ M_\odot$ \citep[cf.][]{bouw12}. However, the
    uncertainties are still large, and therefore the current results are
    also consistent with no change in the sSFR with mass. Using a RSF
    model does not change this result.

  \item At a fixed stellar mass, the derived sSFRs show only a modest
    amount of evolution over the redshift range $z\sim6$ to $z\sim2$ --
    even when the effects of optical emission lines are included. The
    sSFR shows a dependence on redshift ${\rm sSFR \propto
    (1+z)^{1.0\pm0.1}}$, i.e., a factor $\sim2.3\times$ higher at
    $z\sim6$ than at $z\sim2$ (assuming the RSF model with
    \emph{maximal} emission line strength). This appears to be
    inconsistent with simulations, which generally predict a faster
    increase with redshift ${\rm sSFR \propto (1+z)^{2.5}}$, i.e., a
    factor $\sim8.3\times$ higher at $z\sim6$ than at $z\sim2$. The
    evolution we derive is also much slower than observed at lower
    redshifts $z<2$.

  \item The previous conclusion is somewhat in contrast with the
      conclusions derived by \citet{de-b12} and \citet{star13} who argue
      for stronger evolution at $z>2$, similar to that seen in
      simulations. The differences with \citet{star13}, in particular,
      seem to arise from our inclusion of the $M/L$ scatter at a given
      UV luminosity. \citet{star13} also consider the effect of scatter
      in the M/L on the sSFR, but argue that this effect may be offset
      by higher UV luminosity-to-SFR conversion factors which could be
      more common at higher redshift. At present, it is not clear which
      sSFR determination is most accurate, due to the several
      observational uncertainties (e.g., nebular line EWs, intrinsic
      scatter in M/L ratios, see section 5.2). Larger samples, e.g.,
      from the full CANDELS survey, may help elucidate some of these
    currently open questions.

\end{itemize}

\acknowledgements

We acknowledge support from NASA grant HST-GO-11563, and NASA grant
HST-GO-11144. We also acknowledge funding from ERC grant HIGHZ no.
227749. This work is based on observations made with the Hubble Space
Telescope and with the Spitzer Space Telescope. Support for this work
was provided by NASA through an award issued by JPL/Caltech. P.O.
acknowledges support provided by NASA through Hubble Fellowship grant
HF-51278.01 awarded by the Space Telescope Science Institute, which is
operated by the Association of Universities for Research in Astronomy,
Inc., for NASA, under contract NAS 5- 26555.

\end{document}